\documentclass[conference,letterpaper]{IEEEtran}

\IEEEoverridecommandlockouts  %

\usepackage[utf8]{inputenc}
\usepackage[T1]{fontenc}
\usepackage[cmex10]{amsmath}
\usepackage{amssymb}
\usepackage{mathtools}

\usepackage{graphicx}
\usepackage[font=scriptsize]{caption}
\usepackage{subcaption}
\usepackage{booktabs}
\usepackage{makecell}
\usepackage{siunitx}
\usepackage{multirow}

\usepackage{url}
\usepackage{enumitem}
\usepackage{xcolor}
\usepackage{adjustbox}

\usepackage{tikz}
\usetikzlibrary{positioning,trees,shapes,arrows,calc}

\usepackage{hyperref}
\hypersetup{
    colorlinks=true,
    linkcolor=blue,
    filecolor=magenta,
    citecolor=blue,
    urlcolor=blue
}
\usepackage{float}

\newcommand{\figref}[1]{Fig.~\ref{#1}}
\newcommand{\secref}[1]{Sec.~\ref{#1}}
\newcommand{\tabref}[1]{Tab.~\ref{#1}}

\begin{document}

\title{DeepOFDM: Neural Modulation for High Mobility}

\author{
  \IEEEauthorblockN{
  S. Ashwin Hebbar\IEEEauthorrefmark{4}\IEEEauthorrefmark{1},
  Sravan Ankireddy\IEEEauthorrefmark{2}\IEEEauthorrefmark{1},
  Harshithanjani Athi\IEEEauthorrefmark{2},
  Brandon Nguyen\IEEEauthorrefmark{2},
  Pramod Viswanath\IEEEauthorrefmark{4},
  Hyeji Kim\IEEEauthorrefmark{2}
  }
  \IEEEauthorblockA{\IEEEauthorrefmark{2}University of Texas at Austin}
  \IEEEauthorblockA{\IEEEauthorrefmark{4}Princeton University}
  \thanks{\IEEEauthorrefmark{1}Authors contributed equally.}
}

\maketitle
\let\emptyset\varnothing

\begin{abstract}

Orthogonal Frequency Division Multiplexing (OFDM) is the dominant waveform in modern wireless systems, but suffers performance degradation in high-mobility environments due to Doppler-induced inter-carrier interference and unreliable pilot-based channel estimation. Neural receivers have recently shown strong performance in OFDM systems by learning equalization and detection directly from the received time–frequency grid. However, when channel estimation becomes unreliable, receiver-side learning alone is insufficient to fully recover performance.

In this work we introduce DeepOFDM, a learnable modulation framework that augments conventional OFDM with a lightweight convolutional neural network (CNN) modulator jointly optimized with a neural receiver. Instead of mapping symbols independently to resource elements, DeepOFDM spreads information across local time–frequency neighborhoods while remaining fully compatible with FFT-based OFDM processing. The learned modulation breaks the rotational symmetry of conventional QAM constellations, enabling the receiver to infer residual phase directly from data symbols. This structure allows reliable operation with sparse pilots and even in fully pilotless settings. Extensive simulations demonstrate improvements in block error rate and goodput under high Doppler, while over-the-air experiments confirm practical feasibility. These results highlight the potential of transmitter–receiver co-design for robust and spectrally efficient AI-native physical layer design.

\end{abstract}
\section{INTRODUCTION}\label{sec:intro}

\IEEEPARstart{O}{rthogonal} frequency-division multiplexing (OFDM) is the dominant waveform in 4G and 5G networks. It transforms a multipath fading channel into a set of parallel flat-fading subchannels, enabling simple equalization and efficient use of the spectrum. As a result, OFDM delivers strong average-case performance in quasi-static channels, making it well-suited for current 5G deployments.

However, this performance degrades significantly in high-mobility environments, where the wireless channel varies rapidly in both time and frequency. In such settings (e.g., vehicular communications and high-speed rail scenarios), Doppler spread destroys subcarrier orthogonality, causing inter-carrier interference (ICI) and rendering the standard one-tap equalizer ineffective~\cite{ai2014challenges}. At the same time, pilot-based channel estimation becomes increasingly unreliable as channel coherence time shrinks. Under high Doppler, OFDM therefore faces two tightly coupled challenges: equalization error due to ICI and channel inference error due to pilot aging.

These limitations motivate a central question: can OFDM be improved for high-mobility operation without abandoning its underlying structure?
A promising approach is \emph{neural augmentation}, in which analytically designed components of a communication pipeline are replaced or enhanced with learned models while preserving the overall system structure. This paradigm has shown promising results in areas like source and channel coding~\cite{kim2018deepcode,hebbar2022tinyturbo,hebbar2023crisp, ankireddy2023interpreting, li2023neural,hebbar2024deeppolar,ankireddy2025residual}.
Following this principle, recent work has explored neural receivers that replace traditional OFDM receivers, achieving competitive performance without relying on explicit channel estimation~\cite{ye2017power,honkala2021deeprx,wiesmayr2024design}. These data-driven receivers are not restricted to conventional constellations and can adapt to arbitrary modulation patterns, often with improved robustness in time-varying conditions.

In this work, we revisit OFDM modulation under the assumption that a neural receiver is available. Rather than introducing a new waveform, we retain the standard OFDM physical layer and redesign the symbol mapping stage to better align with the inductive biases of convolutional receivers.

Prior transmitter-side learning approaches, including geometric shaping and superimposed pilots ~\cite{stark2019joint, aoudia2021end, madadi2023ai}, improve symbol design but retain the conventional one-symbol-per-RE mapping inherited from classical OFDM. In contrast, we relax this structural constraint.
We propose DeepOFDM, a lightweight convolutional neural modulator that relaxes the strict one-symbol-per-RE constraint while remaining fully compatible with FFT-based processing and cyclic prefix insertion. The modulator learns how to structure transmitted symbols across local time–frequency neighborhoods in a manner tailored to the neural receiver. This joint transmitter–receiver design introduces controlled redundancy and geometric structure that improves robustness when channel estimation becomes challenging.
At the other extreme, fully joint coded modulation approaches learn both coding and modulation end-to-end~\cite{jiang2019turbo, makkuva2021ko, jamali2022productae, hebbar2024deeppolar, ankireddy2025lightcode}, often achieving strong performance but at the cost of scalability and modularity. DeepOFDM occupies a middle ground: it retains standardized LDPC coding and decoding while introducing learnable structure at the modulation stage. This preserves practical system compatibility while capturing many of the robustness benefits of transmitter–receiver co-design.

Our analysis reveals a clear regime-dependent behavior. When channel state information can be accurately inferred, e.g., at low Doppler or with sufficient pilot density, neural receivers alone achieve strong performance, and transmitter learning offers limited benefit. However, as Doppler increases and channel estimation degrades, transmitter-side learning yields substantial gains. These improvements persist and amplify under pilot-sparse configurations. Interestingly, the learned modulation breaks the rotational symmetry inherent in conventional QAM constellations, enabling improved phase identifiability in the presence of Doppler-induced phase uncertainty. As a result, DeepOFDM can operate reliably even in fully pilotless settings, translating into significant goodput improvements. This observation suggests that transmitter-side structure can partially substitute for explicit pilot signaling in rapidly varying channels.

Beyond robustness in specific channel conditions, we show that the learned modulation generalizes across unseen channel models and OFDM grid configurations without retraining, indicating that it captures structural properties of doubly selective channels rather than overfitting to a specific channel model. Finally, we validate the proposed approach through over-the-air experiments using software-defined radios, confirming its practical feasibility in  wireless environments with various impairments.

Preliminary versions of this work appeared in \cite{ankireddy2025ntfofdm, ankireddy2025deep}, where we introduced DeepOFDM for high-mobility OFDM systems. Our main contributions are summarized as follows:
\begin{itemize}

    \item We introduce \emph{DeepOFDM}, a CNN-based neural modulator that relaxes the one-symbol-per-resource-element mapping of conventional OFDM while preserving full compatibility with the OFDM physical layer. This lightweight modulator is jointly optimized with a neural receiver.

    \item We systematically analyze OFDM performance under increasing Doppler. We show that when channel state information can be accurately inferred, neural receivers alone achieve strong equalization performance. However, as Doppler increases and channel estimation becomes unreliable, transmitter-side learning yields substantial gains. These gains persist and amplify under pilot-sparse configurations.

    \item 
    We identify the mechanism underlying these gains: the neural modulator learns an asymmetric constellation structure that breaks the $\pi/2$ rotational symmetry of conventional QAM. Under Doppler-induced phase uncertainty, this asymmetry allows the receiver to infer residual phase directly from the structure of received data symbols, partially substituting for explicit pilot reference. 

    \item We show that DeepOFDM enables fully pilotless operation with negligible degradation relative to pilot-aided settings, translating into significant goodput improvements.

    \item We demonstrate that DeepOFDM generalizes across unseen channel models and OFDM grid configurations without retraining. Furthermore, we validate the approach through over-the-air experiments, confirming practical feasibility in real wireless environments.

\end{itemize}

\section{SYSTEM MODEL AND CONVENTIONAL BASELINES}\label{sec:sysmodel}

\begin{figure*}[!htb]
    \centering
 	\includegraphics[width=\linewidth]{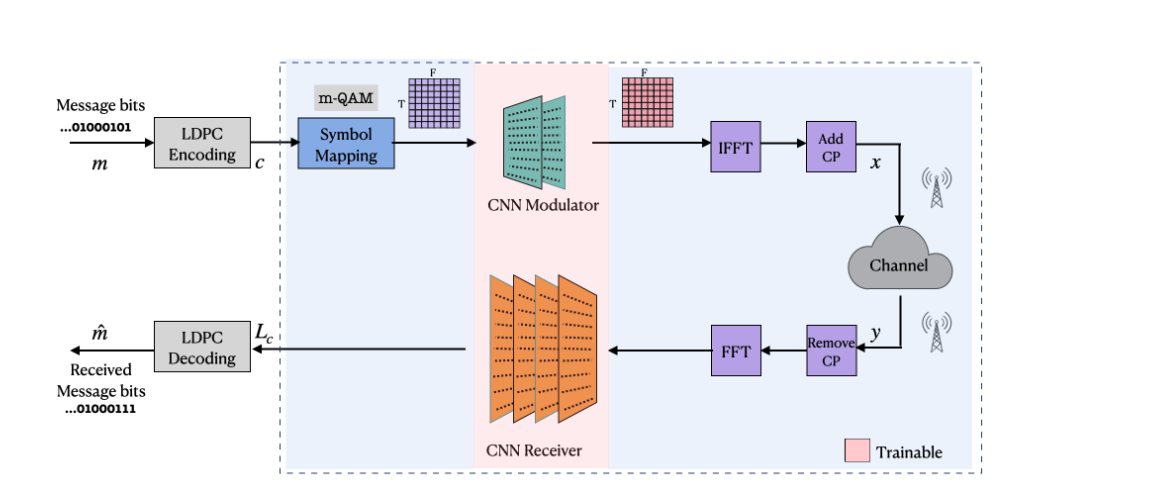}
 	\captionsetup{font=small}
 	\caption{ End-to-end optimization of transmitter and receiver. }
    \label{fig:simple_arch}

\end{figure*}

\subsection{System Model}\label{subsec:sysmodel_detail}
We consider a single-input single-output (SISO) communication system. At the transmitter, the message bit stream $\mathbf{b} \in \{0,1\}^K$, is encoded using a rate $\frac{K}{N}$ LDPC code, producing a codeword $\mathbf{c} \in \{0,1\}^N$. The encoded sequence is mapped to complex symbols using either a standard $2^m$-QAM constellation or a learned constellation, resulting in the modulated symbols $\mathbf{s} \in \mathbb{C}^N$.

These symbols are transmitted over a 5G NR-compliant OFDM system with $n_{S}$ subcarriers and $n_{T} = 14$ OFDM symbols per slot. The frequency-domain symbols are reshaped into a time-frequency grid $\mathbf{X} \in \mathbb{C}^{n_S \times n_T}$ and transformed to the time domain using an inverse FFT, followed by cyclic prefix (CP) insertion.

We consider a time-varying multipath channel modeled by a tapped delay line (TDL)~\cite{3gppTR38901}. 
A user moving at speed $u$~m/s induces a maximum Doppler shift $f_d = \tfrac{u f_c}{c}$, where $f_c$ is the carrier frequency and $c$ is the speed of light.
After pulse shaping, matched filtering, and sampling at period $T_s$ (sample rate $1/T_s$), the discrete-time baseband received signal is
\begin{equation}
\label{eq:tdl-siso}
y[b] \;=\; \sum_{\ell=L_{\min}}^{L_{\max}} h_\ell[b]\; x[b-\ell] \;+\; w[b], \qquad b=0,1,\dots,
\end{equation}
where $x[b]\in\mathbb{C}$ and $y[b]\in\mathbb{C}$ are the transmitted and received samples, 
$w[b]\sim\mathcal{CN}(0,N_0)$ is AWGN, and $h_\ell[b]\in\mathbb{C}$ is the $\ell$-th time-varying tap at sample time $bT_s$.
The tap support $\ell\in\{L_{\min},\dots,L_{\max}\}$ is chosen from the channel's delay spread (e.g., TDL-A profile), with powers $\{p_\ell\}$ satisfying $\sum_\ell p_\ell=1$.

In low-mobility settings, the equivalent frequency domain channel matrix $\mathbf{H}$ is approximately diagonal, but in high-mobility environments, the channel varies within an OFDM symbol duration, breaking the orthogonality between subcarriers. This results in inter-carrier interference (ICI), where $\mathbf{H}$ becomes a full matrix with non-zero off-diagonal elements. The magnitude of these off-diagonal elements increases with mobility, significantly degrading the performance of conventional equalization techniques.

In practice, receivers perform equalization in the frequency domain to keep computational complexity manageable. Accordingly, for frequency-domain equalization we approximate the per–subcarrier/per–OFDM-symbol relationship as
\begin{equation}
\mathbf{Y} = \mathbf{H} \odot \mathbf{X} + \mathbf{W},
\end{equation}
where $\mathbf{X}\in\mathbb{C}^{n_S\times n_T}$ denotes the transmitted time–frequency grid, $\mathbf{H}\in\mathbb{C}^{n_S\times n_T}$ is the effective channel on that grid, $\mathbf{W}\in\mathbb{C}^{n_S\times n_T}$ is additive white Gaussian noise, and $\odot$ indicates the Hadamard (elementwise) product.

In high-mobility settings, the resulting ICI and rapid channel fluctuations violate the assumptions behind conventional pilot-based processing. Next, we review some conventional baselines.

\subsection{Conventional Pilot-Based Receivers}\label{subsec:pilot_rx}
\subsubsection{Least Squares Estimator (LS)}
The LS estimator represents the simplest pilot-based approach  for channel estimation in OFDM systems. Given the transmitted and received pilot symbols $\mathbf{X}_{t,\mathcal{P}}$ and $\mathbf{Y}_{t,\mathcal{P}}$  on the pilot subcarriers $\mathcal{P}_t \subset \{1, \dots, n_S\}$ in the $t$-th OFDM symbol, the diagonal elements of the channel matrix are estimated as
\begin{equation}
    \hat{\mathbf{H}}_{t,P} = \mathrm{diag}\!\left(
        \frac{\mathbf{Y}_{t,\mathcal{P}}}{\mathbf{X}_{t,\mathcal{P}}}
    \right),
    \label{eq:ls_estimate}
\end{equation}
where the division is element-wise. 
Channel estimates for the remaining subcarriers are obtained  through interpolation across $\hat{\mathbf{H}}_{t,\mathcal{P}}$.  While LS estimation performs adequately in low-mobility settings, it suffers from several limitations in high-mobility environments.

\subsubsection{Linear Minimum Mean Square Error Estimator (LMMSE)}
A widely used baseline for OFDM systems with imperfect channel knowledge is the 
LMMSE estimator, which performs channel estimation from orthogonal pilot symbols 
based on the known tempo-spectral covariance matrix \( \mathbf{R} = \mathbf{R_F} \otimes \mathbf{R_T} \), where \(\mathbf{R_F}\) and \(\mathbf{R_T}\) capture frequency and time correlation, respectively. 

We denote by \( \mathbf{P} \in \mathbb{C}^{n_S \times n_T} \) the pilot matrix, 
where \( \mathbf{P}_{i,k} = 0 \) if the resource element (RE) on subcarrier \( i \) 
and time slot \( k \) carries data, and \( \mathbf{P}_{i,k} \) equals the pilot value otherwise. 
Let \( n_P \) denote the number of pilot-bearing REs used for channel estimation. 
Stacking the received samples corresponding to these pilot positions yields 
the vectorized channel model
\begin{equation}
    \mathbf{y}_P = \mathbf{\Pi}\!\left( \mathrm{diag}(\mathbf{p})\,\mathbf{h} + \mathbf{w} \right),
    \label{eq:pilot_obs}
\end{equation}
where \( \mathbf{p} = \mathrm{vec}(\mathbf{P}) \), \( \mathbf{w} = \mathrm{vec}(\mathbf{W}) \),  \( \mathbf{h} = \mathrm{vec}(\mathbf{H}) \)
and \( \mathbf{\Pi}\) is the selection matrix 
that extracts the pilot-bearing REs. 

For the received pilot samples \( \mathbf{y}_P \), 
the LMMSE channel estimate is
\begin{equation}
\begin{aligned}
    \hat{\mathbf{h}} 
    &= \mathbf{R}\,\mathrm{diag}(\mathbf{p})^{H}\mathbf{\Pi}^{H} \\
    &\quad\cdot
    \Big(
        \mathbf{\Pi}\big(
        \mathrm{diag}(\mathbf{p})\mathbf{R}\mathrm{diag}(\mathbf{p})^{H} 
        + \sigma^{2}\mathbf{I}_{n}
        \big)\mathbf{\Pi}^{H}
    \Big)^{-1}\!\mathbf{y}_{P}.
\end{aligned}
\label{eq:lmmse_est}
\end{equation}
and the estimation error correlation matrix is given by
\begin{equation}
\begin{aligned}
    \tilde{\mathbf{R}}
    &= \mathbf{R}
     - \mathbf{R}\,\mathrm{diag}(\mathbf{p})^{H}\mathbf{\Pi}^{H} \\
    &\quad\cdot
    \Big(
        \mathbf{\Pi}\big(
        \mathrm{diag}(\mathbf{p})\mathbf{R}\mathrm{diag}(\mathbf{p})^{H} 
        + \sigma^{2}\mathbf{I}_{n}
        \big)\mathbf{\Pi}^{H}
    \Big)^{-1}
    \mathbf{\Pi}\,\mathrm{diag}(\mathbf{p})\mathbf{R}.
\end{aligned}
\label{eq:lmmse_cov}
\end{equation}

The received signal is modeled as 
\(
\mathbf{y} = \mathrm{diag}(\hat{\mathbf{h}})\mathbf{x} 
            + \mathrm{diag}(\tilde{\mathbf{h}})\mathbf{x} 
            + \mathbf{w},
\)
where \( \tilde{\mathbf{h}} = \mathbf{h} - \hat{\mathbf{h}} \) represents the channel estimation error.
The receiver computes bit-wise log-likelihood ratios (LLRs) for soft decoding assuming  \(\mathrm{diag}(\tilde{\mathbf{h}})\mathbf{x} 
            + \mathbf{w}\) to be Gaussian.
For a modulation order of \( m \) bits per symbol and constellation 
\( \mathcal{C} = \{c_1, \ldots, c_{2^m}\} \), 
let \( \mathcal{C}_{i,0} \) and \( \mathcal{C}_{i,1} \) denote the subsets 
of constellation points whose \( i\)th bit is 0 or 1, respectively. 
For each resource element (RE) \( k \), the LLR of bit \( i \) is given by
\begin{equation}
    \mathrm{LLR}(k,i) = 
    \ln
    \frac{
        \sum\limits_{c\in \mathcal{C}_{i,1}}
        \exp\!\left(-\frac{|y_k - \hat{h}_k c|^2}{\tilde{\sigma}_k^2}\right)
    }{
        \sum\limits_{c\in \mathcal{C}_{i,0}}
        \exp\!\left(-\frac{|y_k - \hat{h}_k c|^2}{\tilde{\sigma}_k^2}\right)
    },
    \label{eq:llr}
\end{equation}
where 
\( \tilde{\sigma}_k^2 = {\tilde{\mathbf{R}}_{k,k}} + \sigma^2 \) 
represents the effective noise variance. LLRs are computed only for data REs and are then passed to the channel decoder (e.g., belief propagation) for soft bit decoding. This non-iterative baseline performs channel estimation and demapping once per frame.

\subsubsection{Iterative LMMSE Estimator (IEDD)}
The iterative estimation, demapping, and decoding (IEDD) estimator
extends LMMSE by using decoder feedback to iteratively refine the channel estimate. 
At each iteration, the decoder provides \emph{a priori} LLRs, 
\( \mathrm{LLR}_P(k,i) \), which are used to compute a prior distribution
\( P_{X_k}(c) \) for each transmitted symbol \( x_k \in \mathcal{C}\) via a softmax mapping. 
Using these priors, the updated LMMSE estimate is computed as
\begin{equation}
    \hat{\mathbf{h}}' 
    = \mathbf{R}\,\mathrm{diag}(\bar{\mathbf{x}})^{H}
    \big(
        \mathbf{R} \circ \mathbb{E}\{\mathbf{x}\mathbf{x}^{H}\} 
        + \sigma^{2}\mathbf{I}
    \big)^{-1} 
    \mathbf{y},
    \label{eq:iedd_est}
\end{equation}
where 
\(
\bar{x}_{k} := \mathbb{E}_{P_{X_{k}}}\{x_{k}\}
\)
and
\[
\mathbb{E}\{\mathbf{x}\mathbf{x}^{H}\}_{i,k} =
\begin{cases}
    \mathbb{E}_{P_{X_{k}}}\{|x_{k}|^{2}\}, & \text{if } i = k, \\[0.75em]
    \mathbb{E}_{P_{X_{i}}}\{x_{i}\}\,
    \mathbb{E}_{P_{X_{k}}}\{x_{k}^{*}\}, & \text{otherwise.}
\end{cases}
\label{eq:iedd_exx}
\]
New extrinsic LLRs are generated as
\begin{equation}
\mathrm{LLR}(k,i) = 
\ln \frac{
\sum_{c \in \mathcal{C}_{i,1}} 
\exp\!\left(-\frac{|y_k - \hat{h}_k c|^2}{\tilde{\sigma}_k'^2} 
+ \sum_l c^{(l)}\mathrm{LLR}_E(k,l)\right)}
{\sum_{c \in \mathcal{C}_{i,0}} 
\exp\!\left(-\frac{|y_k - \hat{h}_k c|^2}{\tilde{\sigma}_k'^2} 
+ \sum_l c^{(l)}\mathrm{LLR}_E(k,l)\right)},
\end{equation}
with \( \tilde{\sigma}_k'^2 = \tilde{\mathbf{R}}'_{k,k} + \sigma^2 \).
Through iterative exchange of soft information between the estimator and decoder, IEDD  effectively performs data-aided channel estimation leveraging symbol priors from decoder feedback alongside pilot symbols.

Because IEDD leverages decoder feedback for data-aided channel estimation, it represents the strongest practical classical baseline and serves as the primary reference for evaluating learning-based gains.

\subsection{Frequency-Domain LMMSE Equalization}\label{subsec:lmmse}

All classical baselines employ one-tap frequency-domain LMMSE equalization on each resource element. For subcarrier $k$ in a given OFDM symbol, the received signal is modeled as
\begin{equation}
y_k = h_k x_k + n_k,
\end{equation}
where $h_k$ denotes the effective channel coefficient and $n_k \sim \mathcal{CN}(0, N_0^{\text{eff}})$ represents effective noise.

Given a channel estimate $\hat{h}_k$, the LMMSE estimate of $x_k$ is
\begin{equation}
\hat{x}_k = \frac{\hat{h}_k^*}{|\hat{h}_k|^2 + N_0^{\text{eff}}}\, y_k,
\end{equation}
and the corresponding post-equalization noise variance passed to the demapper is
\begin{equation}
\tilde{\sigma}_k^2 = \frac{N_0^{\text{eff}}}{|\hat{h}_k|^2 + N_0^{\text{eff}}}.
\end{equation}

Under perfect CSI, $\hat{h}_k = h_k$ and the channel estimation error variance is zero. In high-mobility settings, inter-carrier interference (ICI) introduces additional distortion not captured by the diagonal channel model. We model the effective noise variance as
\begin{equation}
N_0^{\text{eff}} = N_0 + \gamma_{\text{ICI}},
\end{equation}
where $\gamma_{\text{ICI}}$ denotes the ICI power fraction. Assuming a Clarke/Jakes Doppler spectrum and unit signal power,
\begin{equation}
\gamma_{\text{ICI}} = 1 - \operatorname{sinc}^2(\nu_{\max}),
\qquad
\nu_{\max} = \frac{v_{\max} f_c}{c\,\Delta f}.
\end{equation}

This Gaussian approximation of ICI %
~\cite{russell1995interchannel, li2001bounds} provides a consistent and tractable baseline for comparison across all classical receivers.

This modeling ensures that all classical baselines account for Doppler-induced ICI through an increased effective noise variance while retaining computationally efficient one-tap equalization.
Both LMMSE and IEDD serve as strong conventional baselines in modern OFDM research. While these conventional architectures rely on explicit pilot-based channel estimation, 
recent learning-based approaches aim to integrate estimation, detection, and decoding 
within a unified neural framework.

\section{DEEPOFDM: ARCHITECTURE AND TRAINING}\label{sec:neural_mod}

Neural network-based receivers (NRx)~\cite{honkala2021deeprx, balevi2019one, aoudia2021end, pihlajasalo2021deep} have emerged as promising alternatives to traditional signal processing methods. In ~\cite{honkala2021deeprx}, a fully convolutional receiver is introduced that jointly performs channel estimation, equalization, and soft demapping directly from the received time–frequency grid. 
By leveraging the local time–frequency structure and the implicit information in data symbols, DeepRx achieves near-optimal performance across multiple modulation orders and 5G-compliant pilot patterns. A related study in~\cite{pihlajasalo2021deep} proposed a CNN-based physical-layer receiver for extreme mobility, capable of handling severe Doppler spreads and ICI using only sparse pilot patterns. In a complementary direction, the work in~\cite{balevi2019one} developed deep-learning receivers for one-bit OFDM systems, demonstrating that jointly learned estimators and decoders can achieve performance close to unquantized baselines despite coarse quantization. These approaches unify channel estimation, interpolation, and demapping into a single trainable model. The NRx takes the received OFDM frame $\mathbf{Y}$, possibly along with pilots, and directly predicts the transmitted bits $\hat{\mathbf{b}}$. Such receivers have demonstrated strong robustness in dynamic channels, particularly in pilot-sparse scenarios \cite{fischer2022adaptive}. Moreover, recent work has demonstrated the practical feasibility of such systems through the real-time deployment of a 5G standard-compliant NRx, incorporating both hardware and model optimizations~\cite{wiesmayr2024design}.

Building on the idea of a trainable receiver, the transmitter can also be made learnable 
through end-to-end optimization of the constellation geometry and bit labeling~\cite{aoudia2021end}. Here, the constellation points and bit mappings are treated as trainable parameters and are jointly optimized with the NRx to minimize the overall binary cross-entropy loss.
Each learned constellation is centered and power-normalized to ensure zero mean and unit average energy to avoid implicit learning of superimposed pilots or DC offsets. A single constellation trained offline is shown to generalize across a wide range of SNRs, Doppler spreads, and delay profiles, achieving comparable or better BER performance than conventional QAM. While GS improves robustness by adapting constellation geometry to channel statistics, it preserves the conventional one-to-one mapping between symbols and resource elements (REs). The time–frequency structure of OFDM therefore remains unchanged, and each RE continues to carry an independently modulated symbol.

We introduce DeepOFDM, a modulation scheme that relaxes the strict one-symbol-per-RE constraint by allowing a lightweight CNN-based transmitter to jointly encode symbols across local time–frequency neighborhoods.  We leverage the inductive bias of CNN-based neural receivers, which process local neighborhoods of subcarriers and time slots. This joint time–frequency processing enables the transmitted waveform to adapt structurally to channel dynamics while remaining compatible with the OFDM framework.

\subsection{DeepOFDM: Neural Modulator Architecture}\label{sec:arch}
In conventional OFDM, the $n_S \times n_T$ time–frequency resource grid is populated with one modulated symbol per resource element (RE), and a classical receiver processes REs on a per-symbol basis for equalization and decoding. In contrast, neural receivers operate on \emph{groups} of neighboring symbols in the OFDM grid to improve estimation of the transmitted data. Consequently, there is no intrinsic requirement that the transmitter modulate strictly at the per-RE level.

Motivated by this observation, we adopt a CNN-based transmitter that encodes the modulated symbols jointly across the time and frequency axes. CNNs are a natural fit for this purpose, as their spatial structure enables the model to effectively couple nearby time-frequency components, potentially enhancing robustness against mobility-induced impairments, such as inter-carrier interference (ICI). Further, they are parameter-efficient compared to other models such as RNNs, transformers. We term this approach \emph{DeepOFDM} modulation, leveraging deep learning to tailor the modulation scheme for the OFDM waveform.
While we focus on a $1\times 1$ SISO configuration, the architecture naturally extends to MIMO by treating antennas as an additional dimension and coding across the resulting 3D time–frequency–antenna grid rather than the 2D grid.

\begin{figure}[!htb]
    \centering
    \begin{subfigure}[b]{0.8\linewidth}
        \centering
        \includegraphics[width=\linewidth]{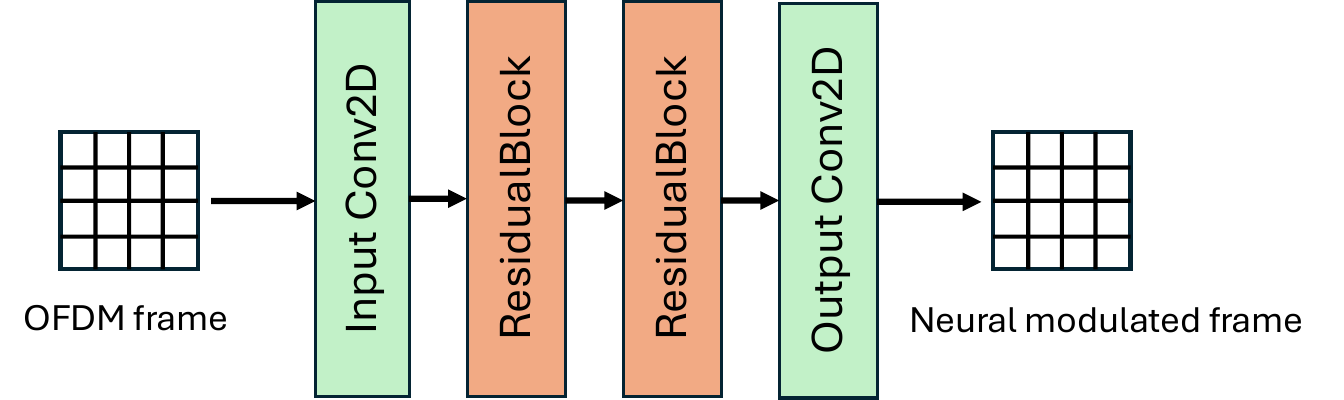}
        \captionsetup{font=small}
        \caption{Neural modulation of OFDM frame}
        \label{fig:neural_mod}
    \end{subfigure}
    
    \vspace{0.5em} %

    \begin{subfigure}[b]{0.8\linewidth}
        \centering
        \includegraphics[width=\linewidth]{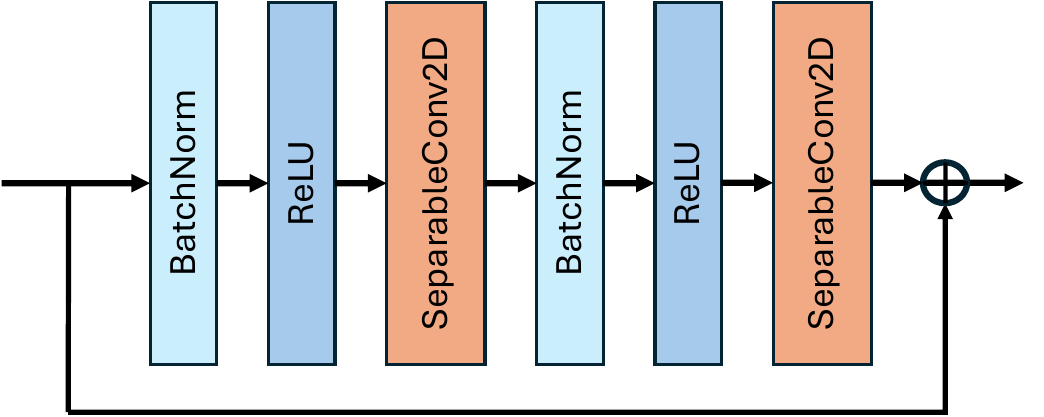}
        \captionsetup{font=small}
        \caption{Parameter efficient \texttt{ResidualBlock}.}
        \label{fig:res_block}
    \end{subfigure}
    
    \captionsetup{font=small}
    \caption{Neural Modulation architecture}
    \label{fig:neural_mod_full}
\end{figure}

Our focus is on improving the transmitter design, which typically operates under tighter power and compute budgets than the receiver. Accordingly, the neural modulator in~\figref{fig:neural_mod_full} is deliberately lightweight, with its parameter budget being constrained to approximately $10\%$ of the receiver, thereby minimizing transmitter-side computational load (see \secref{sec:complexity} for sensitivity analysis of this allocation), the details of which are provided in \tabref{tab:mod} and Table~\ref{tab:nrx}.

As illustrated in~\figref{fig:neural_mod_full}, the modulator ingests complex-valued symbols produced by a learnable constellation mapping applied to encoded bits, arranged as a time–frequency grid $X$. Following~\cite{aoudia2021end}, the constellation is optimized jointly within the end-to-end training. The modulator outputs neural-modulated data $X_m$ in the same time–frequency format, which is subsequently passed through the inverse fast Fourier transform (IFFT) to synthesize the transmit waveform. Finally, a neural receiver is used to decode the received waveform.

\subsection{DeepOFDM: Neural Receiver Architecture}\label{subsec:nrx}
\begin{figure}[!htb]
    \centering
 	\includegraphics[width=0.95\linewidth]{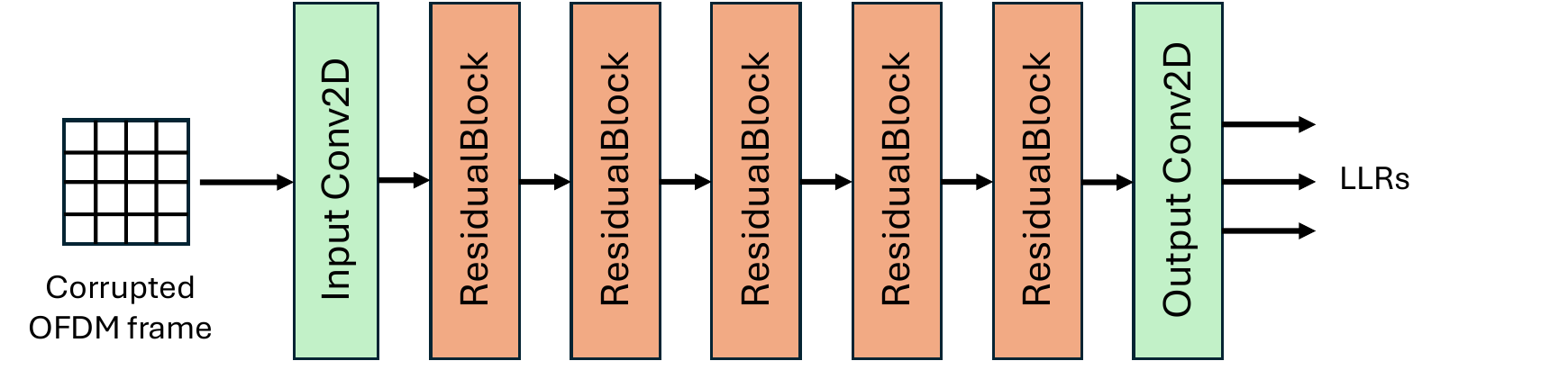}
 	\captionsetup{font=small}
 	\caption{Neural Receiver architecture.}
 	\label{fig:neural_rx}
\end{figure}
We adopt a CNN-based architecture for the NRx, closely following the design introduced in~\cite{aoudia2021end}. The core building block is a \texttt{ResidualBlock}, which contains two convolutional layers. To enhance parameter efficiency without compromising performance, we replace the standard \texttt{Conv2D} layers with \texttt{SeparableConv2D}, as illustrated in~\figref{fig:res_block}. The overall architecture comprises an input convolution layer, followed by five residual blocks, and an output layer that directly predicts the log-likelihood ratios (LLRs) of the transmitted bits, as shown in ~\figref{fig:neural_rx}. The input to the NRx consists of the received OFDM frame (i.e., the complex-valued time-frequency grid after FFT and CP removal), along with the positions and values of any known pilot symbols if present. This setup enables the receiver to flexibly operate in pilot-rich, sparse-pilot, or fully pilotless scenarios.
In contrast to traditional receivers, which employ separate modules for channel estimation, equalization, and demapping, the NRx performs all these tasks jointly through a single unified neural module. This end-to-end formulation unifies the traditionally disjoint processing stages into a single learnable module, maintaining adaptability across diverse channel and pilot configurations. This joint formulation enables the discovery of transceiver strategies that depart from analytically designed OFDM waveforms.

\begin{table}[!htb]
  \centering
\captionsetup{font=small}
\caption{Architecture details for the DeepOFDM modulator.}
\label{tab:mod}
\begin{tabular}{l@{\hskip 4pt}c@{\hskip 4pt}c@{\hskip 4pt}c}
    \toprule
    Layer
      & \makecell[c]{Channels}
      & \makecell[c]{Kernel \\ size}
      & \makecell[c]{Dilation \\ rate} \\
    \midrule
    Input Conv2D      & 24  & (3,3) & (1,1) \\
    ResNet block 1    & 48  & (5,5) & (1,1) \\
    Projection Conv2D      & 48  & (1,1) & (1,1) \\
    ResNet block 2    & 48  & (5,5) & (1,1) \\
    Output Conv2D     & \(2\) & (1,1) & (1,1) \\
    \bottomrule
\end{tabular}
\end{table}

\begin{table}[!htb]
  \centering
\captionsetup{font=small}
\caption{Architecture details for the neural receiver.}
\label{tab:nrx}
\begin{tabular}{l@{\hskip 4pt}c@{\hskip 4pt}c@{\hskip 4pt}c}
    \toprule
    Layer
      & \makecell[c]{Channels}
      & \makecell[c]{Kernel \\ size}
      & \makecell[c]{Dilation \\ rate} \\
    \midrule
    Input Conv2D      & 64   & (3,3) & (1,1) \\
    ResNet block 1    & 128  & (7,5) & (7,2) \\
    ResNet block 2    & 128  & (7,5) & (7,1) \\
    ResNet block 3    & 128  & (5,3) & (1,2) \\
    ResNet block 4    & 128  & (5,3) & (1,2) \\
    ResNet block 5    & 128  & (3,3) & (1,1) \\
    Output Conv2D     & \(2m\) & (1,1) & (1,1) \\
    \bottomrule
\end{tabular}
\end{table}

\begin{table}[h]
\centering
\captionsetup{font=small}
\caption{Training configuration.}
\label{tab:params}
\begin{tabular}{@{}lcc@{}}
\toprule
\textbf{Parameter} & \textbf{Symbol (if any)} & \textbf{Value} \\
\midrule
Number of OFDM symbols     & $n_T$       & 14 (1 slot)        \\
Number of subcarriers      & $n_S$       & 128                \\
Carrier frequency          & --          & 2 {GHz}      \\
Subcarrier spacing         & --          & 15 {kHz}       \\
Cyclic prefix duration     & $n_{CP}$    & 6 symbols          \\
Channel model              & --          & TDL-A              \\
Learning rate              & --          & $10^{-3}$          \\
Batch size for training    & $S$         & 128                \\
Modulation order           & $m$         & 6                  \\
UE speed range (training)  & --          & \SIrange{0}{100}{m/s} \\
\bottomrule
\end{tabular}
\end{table}

\subsection{Training}
We jointly optimize the neural modulator and neural receiver by minimizing an end-to-end loss function. Our objective is to maximize the mutual information between the transmitted bit sequence and the received log-likelihood ratios (LLRs). However, direct maximization of mutual information is generally intractable in practical channel settings. As a practical surrogate, we minimize the binary cross-entropy (BCE) loss between the transmitted bits and the predicted LLRs at the receiver output.

We express this surrogate in terms of the effective rate, and define the loss as
\begin{equation}
\mathcal{L}{\mathrm{rate}} = -\left(1 - \frac{\mathcal{L}_{\mathrm{BCE}}}{\ln 2}\right) = \frac{\mathcal{L}_{\mathrm{BCE}}}{\ln 2} - 1.
\label{eq:loss}
\end{equation} 
This formulation measures the deviation from ideal bitwise mutual information (i.e., 1 bit per binary symbol), and is minimized when the predicted LLRs perfectly match the transmitted bits.

\section{PERFORMANCE EVALUATION}\label{sec:results}

We evaluate DeepOFDM under high-mobility doubly selective channels and compare against both classical and learning-based baselines.

\subsection{Simulation Environment}\label{subsec:sim_env}
All simulations follow the system model described in \secref{sec:sysmodel}, and are implemented using the GPU-accelerated Sionna framework \cite{hoydis2022sionna}. 
The channel model is 3GPP TDL-A with mobility uniformly sampled during training from 0 to 100 m/s (223 mph), covering stationary, vehicular, and high-speed scenarios. Key training parameters are outlined in \tabref{tab:params}.

To reduce training complexity, LDPC encoding and decoding are omitted during training and reintroduced during evaluation with a rate-1/2 LDPC code, whose blocklength equals the number of data-bearing resource elements in the OFDM grid. Unless otherwise stated, models are trained jointly across SNRs and mobility conditions.

We report performance at three representative mobility regimes: 10 m/s (low), 40 m/s (moderate), and 100 m/s (high). These regimes progressively increase Doppler spread and reduce coherence time, making channel tracking more challenging.

To study robustness under varying levels of explicit channel reference information, we evaluate three pilot configurations: (i) two pilot symbols (2P) inserted at the third and twelfth OFDM symbols, (ii) a single pilot symbol (1P) inserted at the third OFDM symbol, and (iii) a fully pilotless setting (0P).

We compare against classical pilot-aided receivers described in \secref{sec:sysmodel}, including LMMSE channel estimation and iterative estimation with decoder feedback (IEDD), both followed by one-tap frequency-domain MMSE equalization. We also evaluate learning-based baselines from \secref{sec:arch}, namely standard QAM modulation with a neural receiver (QAM+NRx) and geometric shaping jointly optimized with the neural receiver (GS+NRx), which maintain a one-to-one symbol-to-resource-element mapping. To ensure a fair comparison, we allocate a near-identical total parameter budget to both DeepOFDM (modulator+receiver) and QAM+NRx.
In addition, we consider an oracle baseline in which the receiver is provided the exact per-subcarrier frequency-domain channel coefficients and performs one-tap MMSE equalization (Freq. domain CSI), thereby isolating performance under perfect diagonal channel knowledge.

\subsection{Performance with Perfect Channel Knowledge}

To better understand the role of transmitter-side learning, we first consider a setting in which channel state information (CSI) is accurately known. This scenario is useful because it decouples the two main challenges of communication over time-varying channels: channel estimation and signal equalization. When CSI is available with high accuracy, the receiver’s task reduces to equalizing the received signal under Doppler-induced interference, allowing us to study equalization performance independently of channel estimation.

In this setting, the exact per-subcarrier frequency-domain channel coefficients $H_{k,t}$ are provided to the receiver. For the neural baselines (QAM+NRx and DeepOFDM), this CSI is concatenated to the received time–frequency grid as an additional input dimension, eliminating the need for channel estimation. The task therefore reduces purely to detection under perfect diagonal channel knowledge.

\begin{figure*}[!t]
  \centering
    \includegraphics[width=0.66\textwidth]{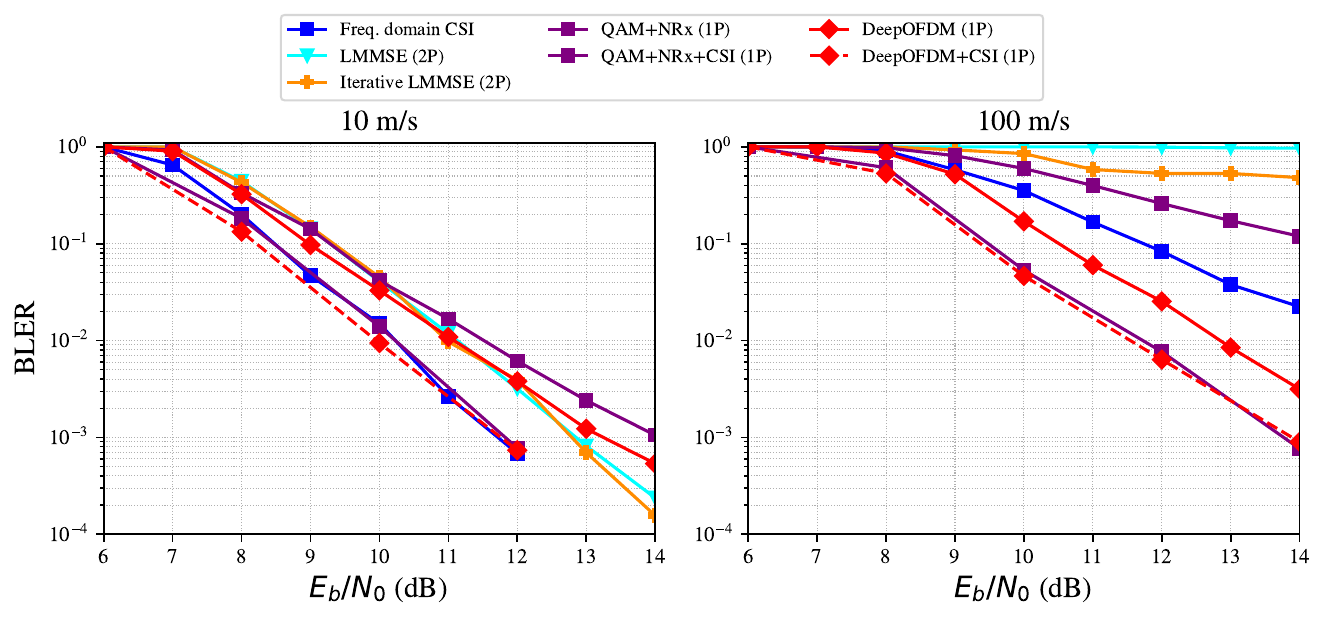}
  \captionsetup{font=small}
    \caption{BLER vs. SNR under perfect frequency-domain CSI. Under accurate channel knowledge, QAM+NRx and DeepOFDM achieve nearly identical performance, confirming that transmitter-side learning provides no benefit when channel estimation is not the bottleneck.}
    \label{fig:rx_csi}
\end{figure*}

\figref{fig:rx_csi} shows the resulting BLER performance. Under perfect CSI, QAM+NRx and DeepOFDM achieve nearly identical performance across all mobility regimes. Both neural receivers also outperform the classical one-tap MMSE equalizer, particularly under higher Doppler effects.

This result makes two points clear. First, neural receivers are already highly effective when accurate channel knowledge is available. Second, in this regime, transmitter-side learning does not provide additional benefit. Any performance differences that emerge in practical settings must therefore be attributed to how well the channel is estimated, rather than how well it is equalized.

We now move to realistic pilot-aided operation, where CSI must be estimated rather than provided.

\subsection{Performance at High Mobility}

\begin{figure*}[ht]
  \centering
  \includegraphics[width=\textwidth]{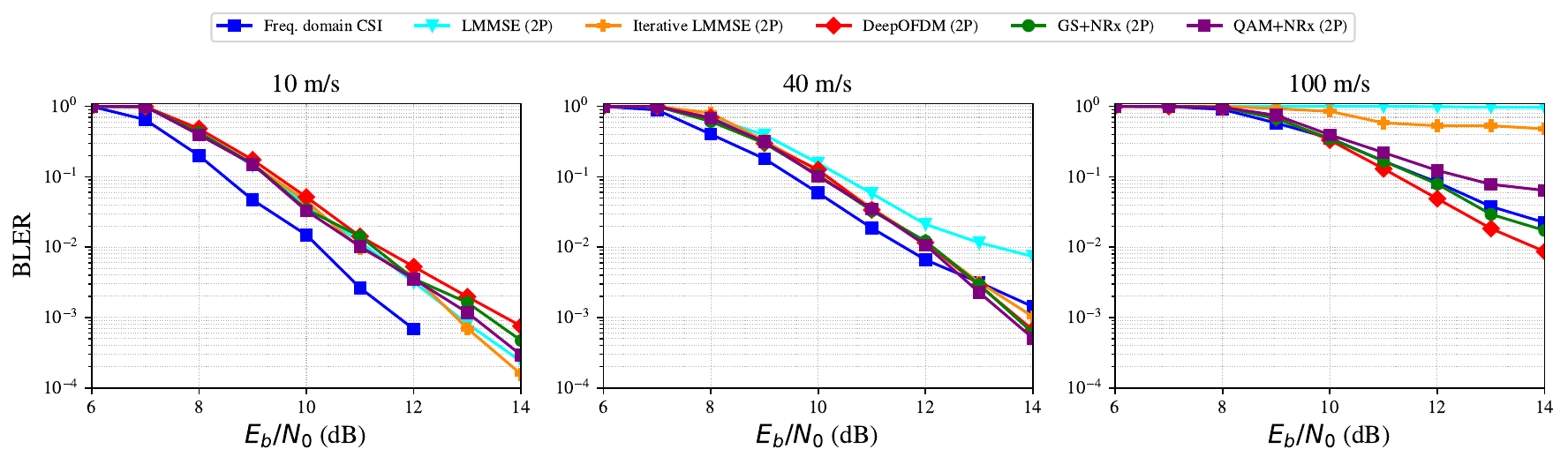}
  \captionsetup{font=small}
  \caption{BLER vs.\ SNR under the 2P configuration (two pilot symbols per slot) on the TDL-A channel. With sufficient pilot reference, learning-based schemes exhibit similar performance at low and moderate mobility (10, 40~m/s), while DeepOFDM provides clear gains at high mobility (100~m/s).}
  \label{fig:2p}
\end{figure*}

We now consider practical pilot-aided operation, where channel state information must be estimated rather than provided. In this setting, CSI is inferred from a finite number of pilot symbols and is therefore imperfect, particularly under high mobility where the channel evolves rapidly within a slot. We adopt a baseline configuration with two pilot symbols per slot (2P), inserted at the third and twelfth OFDM symbols. \figref{fig:2p} shows the resulting BLER across mobility regimes.

At low and moderate mobility (10 m/s and 40 m/s), QAM+NRx, GS+NRx, and DeepOFDM achieve similar performance. In this regime, pilot information is sufficient for reliable channel tracking, and the dominant challenge is equalization rather than channel inference. The neural receiver already compensates for residual distortion beyond one-tap equalization, leaving limited room for additional transmitter-side gains.

The behavior changes markedly at 100 m/s. As channel variation within a slot becomes significant, classical pilot-aided receivers degrade sharply due to intra-symbol channel evolution and inter-carrier interference. Replacing the classical pipeline with a neural receiver recovers a substantial portion of this loss, confirming that learned equalization improves robustness under Doppler. Learning the constellation (GS+NRx) yields further gains by better matching symbol geometry to the neural receiver. DeepOFDM achieves the lowest BLER among the learning-based schemes at 100 m/s, providing approximately 1dB gain over GS+NRx at BLER of $10^{-2}$, with consistent improvement across the full SNR range shown.

This trend suggests that transmitter-side learning provides increasing robustness as the channel varies more rapidly within a slot. By introducing controlled time–frequency coupling, DeepOFDM provides additional robustness beyond what can be achieved by receiver learning alone. We next consider a more challenging regime in which the number of available pilots is further reduced.

\subsection{Performance Under Reduced Piloting}

\begin{figure*}[ht]
  \centering
  \includegraphics[width=\textwidth]{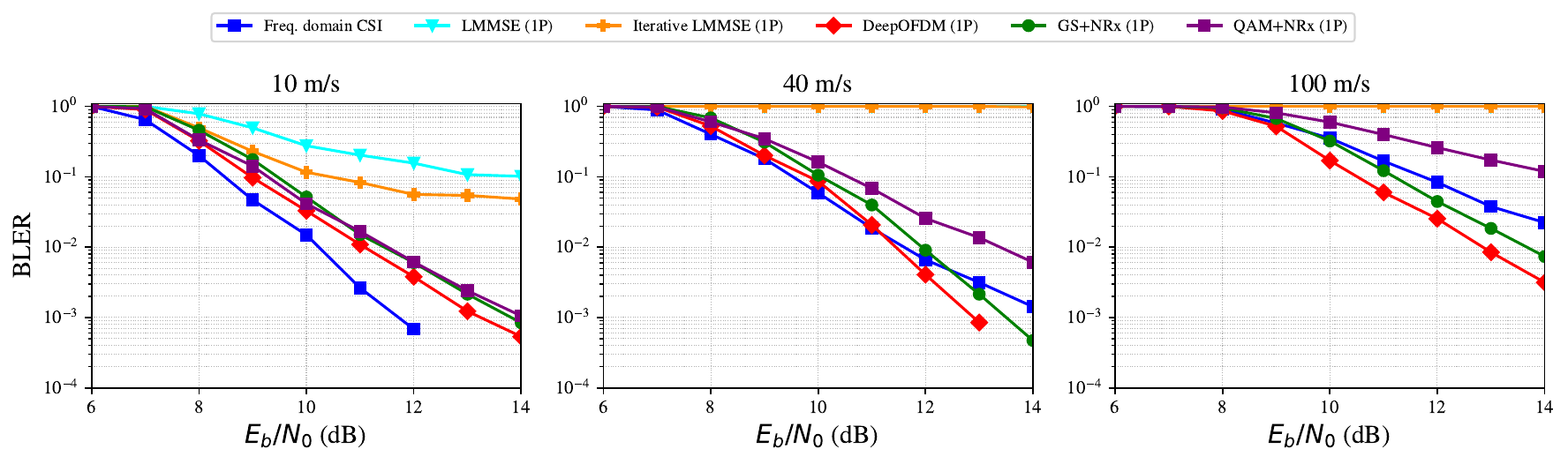}
  \captionsetup{font=small}
  \caption{BLER vs.\ SNR under the 1P configuration on the TDL-A channel. Reducing pilot density exposes clear performance separation, with DeepOFDM maintaining improved reliability at moderate and high mobility.}
  \label{fig:1p}
\end{figure*}

We now consider a more constrained setting with only a single pilot symbol per slot (1P). Compared to the 2P case, this substantially limits the receiver’s ability to track channel variation within a slot. \figref{fig:1p} shows the resulting BLER across mobility regimes.

Reducing pilot density fundamentally changes the operating regime. While the 2P results indicated that performance at moderate mobility (40~m/s) was primarily limited by equalization, the 1P configuration shifts the dominant challenge to channel estimation. Even at 40~m/s, classical pilot-based receivers deteriorate sharply, reflecting the increased difficulty of interpolating a rapidly time-varying channel from a single pilot symbol.

Introducing a neural receiver again improves reliability relative to classical methods, as it can exploit local time–frequency structure beyond explicit pilot locations. Geometric shaping (GS+NRx) provides additional gains over standard QAM, consistent with prior work demonstrating the benefits of constellation optimization under fading. However, the difference in performance between schemes is now clearly visible even at moderate mobility.

DeepOFDM maintains substantially improved reliability across both 40 m/s and 100 m/s. In contrast to the 2P setting, where clear gains emerged mainly at very high mobility, the reduced pilot configuration exposes performance differences even at moderate speeds. At 40m/s and 100 m/s, DeepOFDM provides approximately 0.6~dB and 0.9~dB gain respectively over GS+NRx at a target BLER of $10^{-2}$.

Importantly, all learning-based schemes operate under the same total parameter budget. Thus, the observed gains do not arise from increased model capacity, but from how that capacity is allocated. This highlights a key architectural insight: jointly optimizing transmitter and receiver yields better robustness than concentrating model capacity entirely at the receiver. 

Taken together, these results indicate that as explicit channel reference becomes more limited, transmitter-side structure becomes increasingly valuable. We next examine the structural properties of the learned modulation that enable this behavior.

Finally, we note that the performance trends observed in the sparse-pilot regime suggest that DeepOFDM relies less heavily on explicit pilot signaling than conventional OFDM. We analyze this behavior in detail in the next section.

\section{INTERPRETATION AND ANALYSIS}\label{sec:interpretation}

The results in \secref{sec:results} indicate that DeepOFDM maintains reliability under reduced pilot reference and high mobility. In this section, we examine how the geometry of the learned modulation contributes to this behavior.

\subsection{Constellation Geometry Under Doppler}

Under additive white Gaussian noise (AWGN), square QAM constellations are near-optimal in the sense of maximizing minimum Euclidean distance for a given average power. Their rotational symmetry does not pose a problem when the receiver has precise phase information.

In high-mobility channels, however, pilot-based interpolation cannot perfectly track the channel within an OFDM slot. As a result, a residual phase mismatch remains. The effective detection model can be approximated as
\[
y = e^{j\theta} x + w,
\]
where $\theta$ represents a small but non-negligible phase error.

For QAM, the constellation satisfies rotational symmetry under multiples of $\pi/2$, i.e.,
\[
\mathcal{X} = e^{j\pi/2} \mathcal{X}.
\]
Consequently, a rotated symbol cloud remains geometrically consistent with the original constellation up to $\pi/2$ rotations. In the absence of strong pilot reference, the residual phase is therefore only identifiable modulo $\pi/2$ from the symbol geometry alone.

This suggests that, under phase uncertainty induced by Doppler, constellations lacking rotational symmetry may provide improved phase identifiability from data symbols themselves. 
We now verify that the neural modulator indeed learns this asymmetric structure after end-to-end training.

\subsection{Learned Constellation Structure Under High Mobility}

\begin{figure}[h]
    \centering
    \includegraphics[width=0.7\linewidth]{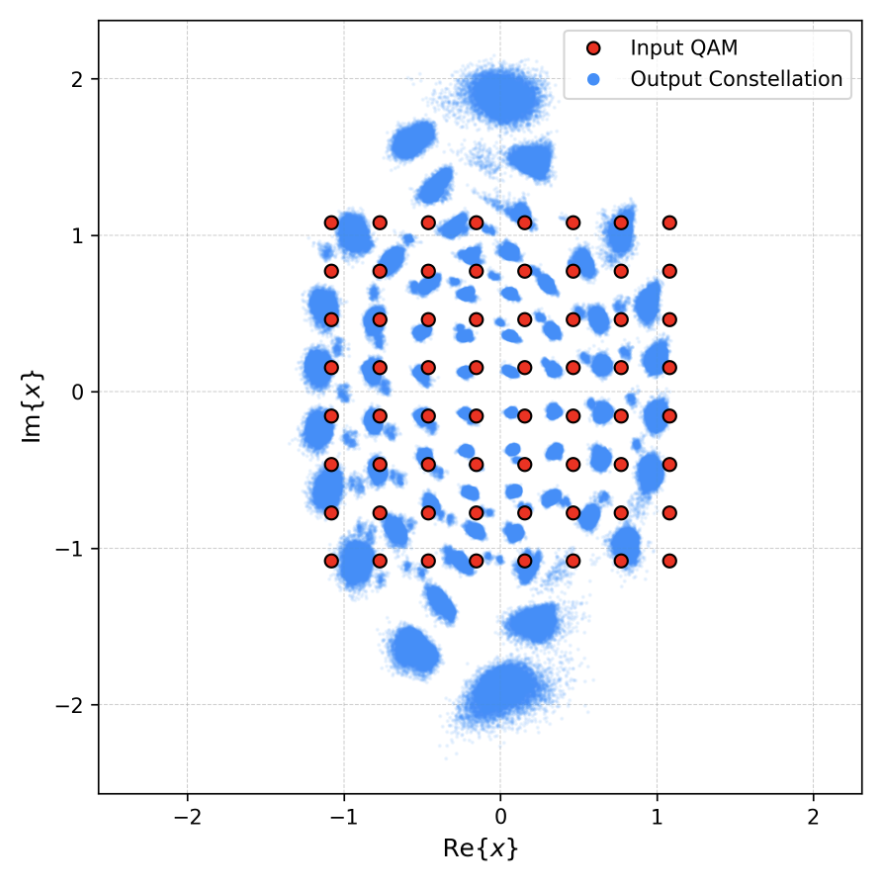}
    \captionsetup{font=small}
    \caption{The output of the neural modulator displays asymmetry -this property enables reliable communication in the absence of pilots}
    \label{fig:asymeetric}
    \vspace{-1em}
\end{figure}

We now examine the constellation learned by the neural modulator after end-to-end training under high-mobility conditions.

\figref{fig:asymeetric} shows the learned outputs overlaid with the original QAM grid. Unlike square QAM, the learned constellation lacks rotational invariance. The symbol geometry is clearly asymmetric and exhibits uneven angular structure. In contrast to conventional modulation, the mapping is not restricted to a single constellation point per symbol; instead, due to the time-frequency mixing introduced by the neural modulator, each input symbol maps to a set of nearby output points rather than a single deterministic constellation location.

This learned asymmetry aligns with the design insight of the previous subsection, and allows phase to be inferred from data symbols themselves.

This behavior is consistent with the empirical trends observed in \secref{sec:results}: when channel knowledge is accurate, performance differences are negligible; as pilot information becomes sparse or outdated, the relative advantage of DeepOFDM increases.

\subsection{Ablation Study: Role of Symbol Diversity}

\begin{figure}[t]
    \centering
    \includegraphics[width=0.9\linewidth]{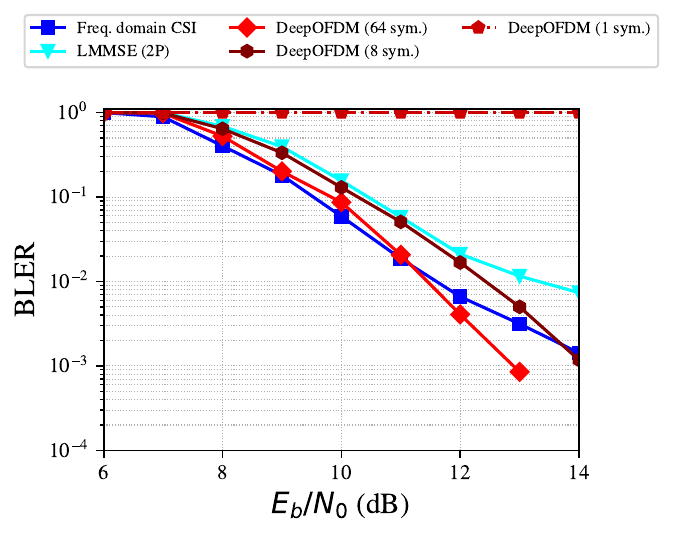}
    \captionsetup{font=small}
    \caption{BLER vs.\ SNR under 1P pilots at 40 m/s. Collapsing the constellation to a single symbol causes catastrophic degradation of DeepOFDM, while restricting to eight symbols restores reliable performance.}
    \label{fig:geometry_ablation}
    \vspace{-1em}
\end{figure}

To isolate the role of constellation geometry, we perform a controlled ablation study in which the diversity of transmitted data symbols is artificially restricted while keeping the transmitter, receiver, and the 1P pilot configuration fixed. The resulting BLER performance is shown in \figref{fig:geometry_ablation}.

\textbf{Single-Symbol Configuration:}  
In the first experiment, all data-bearing resource elements within each codeword are forced to carry the same complex symbol. Instead of mapping iid bits to multiple constellation points, a single fixed constellation point is transmitted across the entire time–frequency grid.

In conventional coded OFDM systems, channel coding and scrambling ensure that transmitted symbols are approximately independent and uniformly distributed over the constellation. This diversity stabilizes signal power, preserves balanced real–imaginary statistics, and matches the assumptions under which the receivers are trained. When all data symbols are identical, this statistical structure collapses.

For QAM+NRx, the resulting degradation is moderate and primarily attributable to the loss of symbol-level diversity, which affects normalization and the statistical regularities exploited by the NRx. However, detection remains functional.

In contrast, DeepOFDM degrades catastrophically: BLER approaches one across the evaluated SNR range. In this configuration, the effective constellation collapses to a single point, eliminating any geometric asymmetry across the grid. As a result, the data symbols no longer provide orientation information that can assist in stabilizing residual phase mismatch. This behavior indicates that DeepOFDM has learned to rely critically on constellation asymmetry for robust operation, and does not behave like a conventional pilot-anchored receiver.

\textbf{Restricted-Symbol Configuration:}  
We next repeat the experiment using a restricted alphabet of eight distinct constellation points selected from the learned DeepOFDM constellation. These symbols are assigned randomly across the grid, preserving spatial and temporal variation while reducing the number of distinct constellation points.

Under this configuration, DeepOFDM restores reliable operation, with BLER significantly improved relative to the single-symbol case. This indicates that even limited symbol diversity is sufficient to recover the asymmetric geometric structure necessary for robust phase tracking.

These experiments confirm that the performance gains observed in Sections~IV-C and IV-D arise primarily from the lack of rotational invariance in the learned output symbols.

\section{PILOTLESS COMMUNICATION}

\begin{figure}[h]
    \centering
    \includegraphics[width=0.9\linewidth]{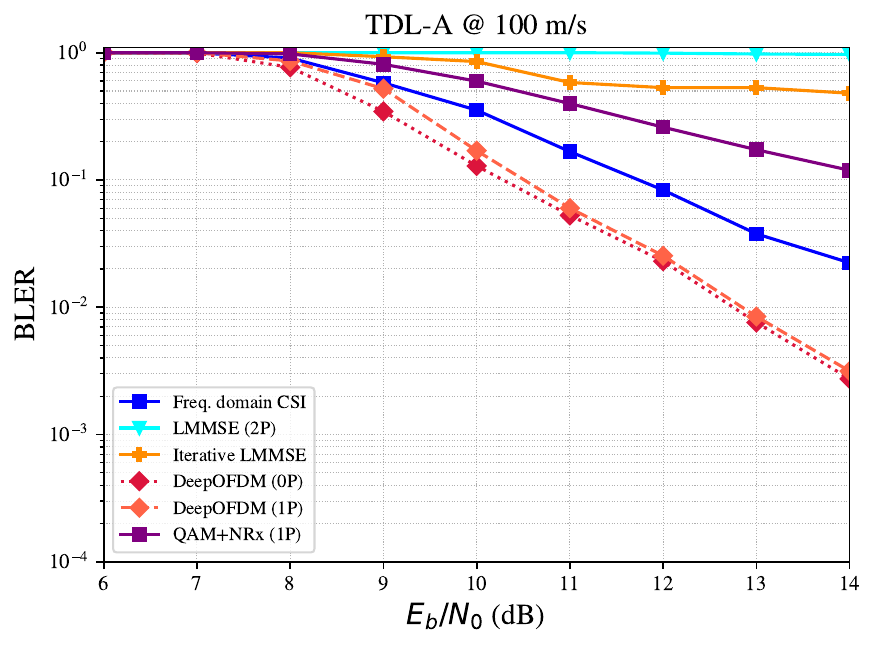}
    \captionsetup{font=small}
    \caption{BLER vs.\ SNR comparing DeepOFDM with a single pilot (1P) and no pilots (0P).
    DeepOFDM exhibits negligible performance degradation when operating without pilots, demonstrating reliable pilotless communication.}
    \label{fig:0p_v_1p}
    \vspace{-1em}
\end{figure}

The analysis in \secref{sec:interpretation} showed that the outputs of DeepOFDM break rotational symmetry, making residual phase identifiable from the received symbol geometry. This raises a natural and practically important question: if data symbols themselves provide sufficient structure for channel inference, are explicit pilot symbols strictly necessary?

We now evaluate DeepOFDM in a fully pilotless configuration (0P), where no time-frequency resources are reserved for pilot symbols.
Pilot-free operation is highly desirable in short-packet, ultra-low-latency, or bandwidth-constrained settings, since orthogonal pilots directly reduce spectral efficiency.
However, removing pilots fundamentally breaks the assumptions of conventional coherent OFDM receivers, which rely on explicit reference signals for channel estimation and synchronization.

\subsection{From Sparse Pilots to No Pilots}

\begin{figure}[h]
    \centering
    \includegraphics[width=\linewidth]{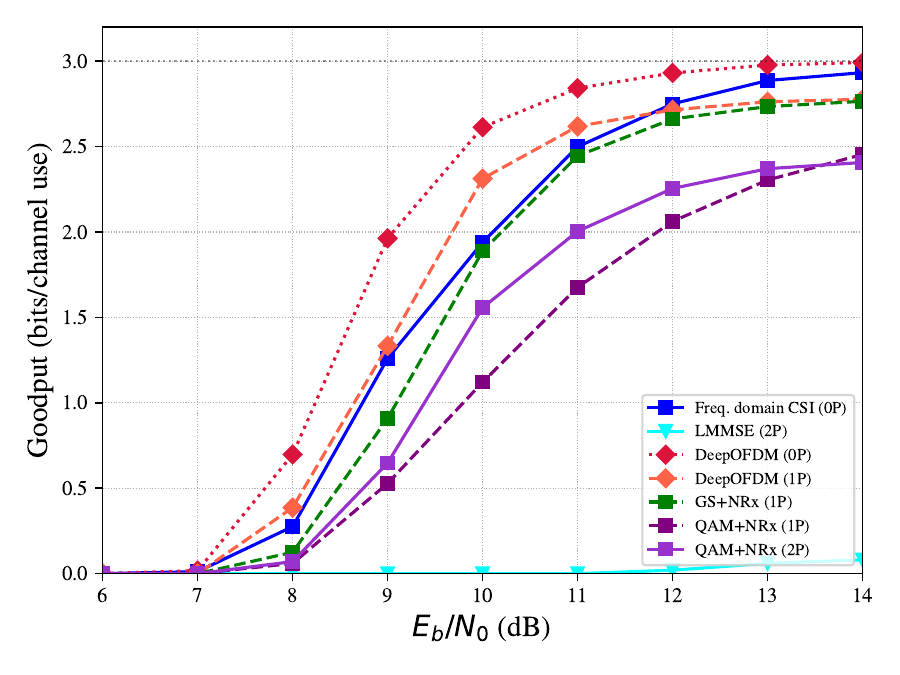}
    \captionsetup{font=small}
    \caption{Goodput vs.\ $E_b/N_0$ at 100~m/s under 0P and pilot-based configurations. DeepOFDM achieves higher effective throughput by eliminating pilot overhead while maintaining low BLER.}
    \label{fig:goodput}
    \vspace{-1em}
\end{figure}

\begin{figure*}[ht]
  \centering
  \includegraphics[width=\textwidth]{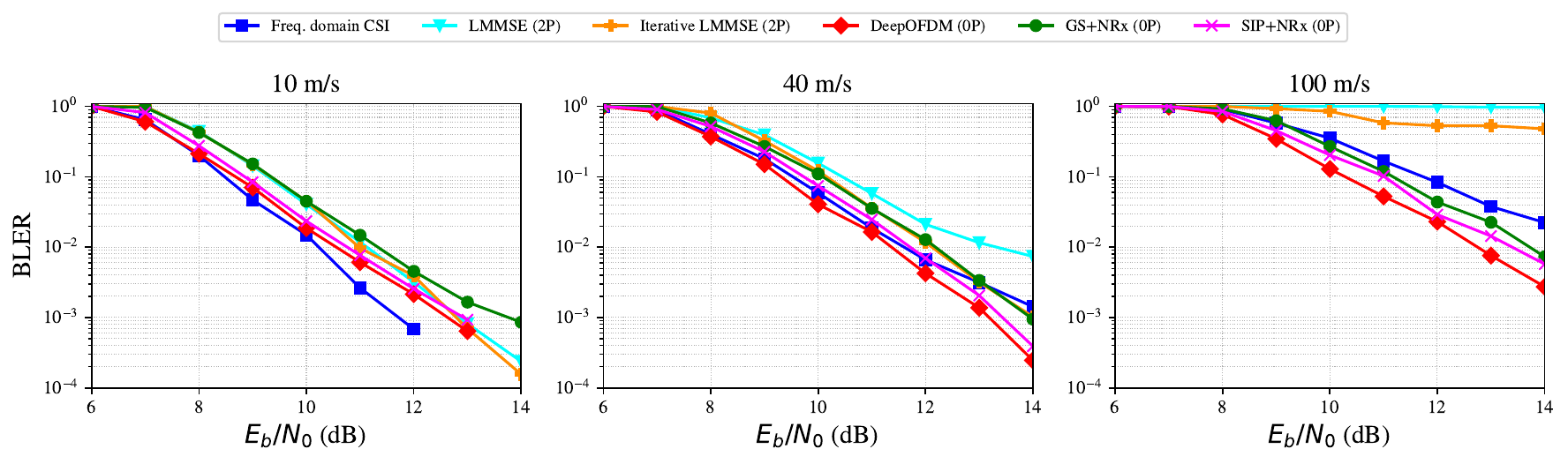}
  \captionsetup{font=small}
  \caption{BLER vs SNR on TDL-A channel. Pilot-free DeepOFDM + NRx attains the lowest BLER across speeds, outperforming other pilot-free neural baselines and classical methods that use 2 pilots.}
  \label{fig:0p}
\end{figure*}

\begin{figure*}[t]
  \centering
  \includegraphics[width=\textwidth]{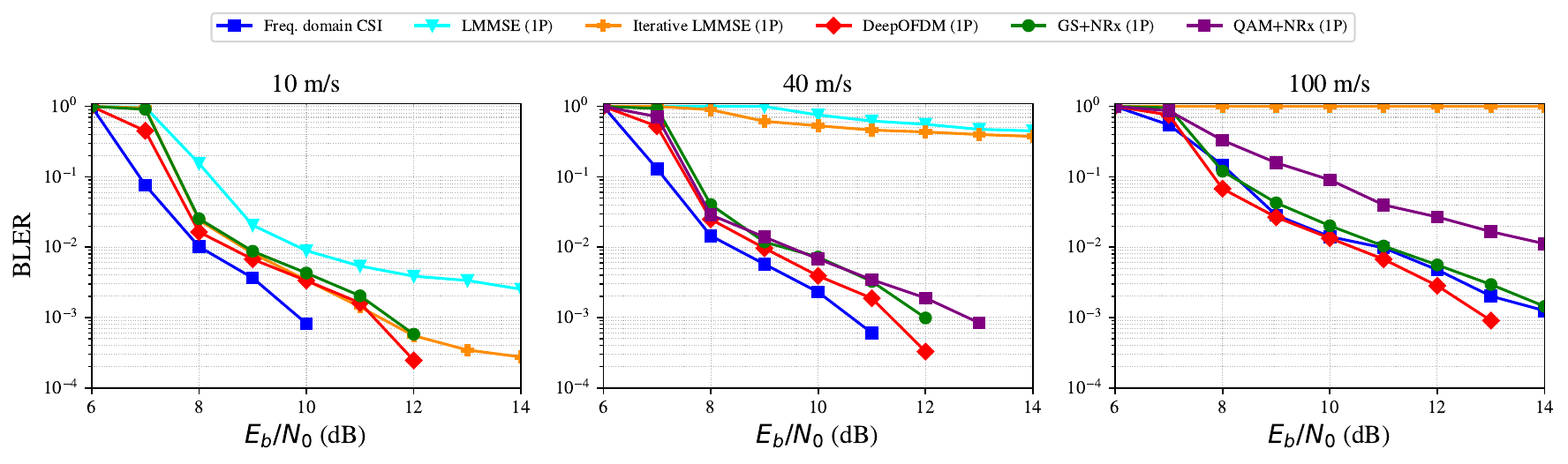}
  \captionsetup{font=small}
  \caption{BLER vs SNR on CDL-C channel, using models trained on TDL-A channel. DeepOFDM is robust to channel models that are unseen in the training phase.}
  \label{fig:cdl}
\end{figure*}

\figref{fig:0p_v_1p} compares DeepOFDM under the single-pilot (1P) and pilotless (0P) configurations.
Remarkably, DeepOFDM exhibits almost no degradation when transitioning from 1P to 0P, demonstrating that DeepOFDM does not require explicit pilot resources to maintain reliable decoding. In contrast, conventional OFDM pipelines fail without pilots, as least-squares and LMMSE estimators are ill-posed in the absence of reference symbols.
Similarly, learned receivers designed around explicitly labeled pilot locations such as QAM+NRx degrade substantially under pilotless setting, since the receiver architecture and training distribution implicitly assume pilot-provided reference.

\subsection{DeepOFDM achieves significant Goodput gains.}
Beyond reliability, pilot-sparse and pilotless operation enables a direct improvement in effective throughput.
We quantify this using the achieved goodput,
\begin{equation}
    \text{Goodput} = R \cdot \rho \cdot (1 - \text{BLER}),
\end{equation}
where $R$ denotes the transmission rate in bits per channel use and $\rho = 1 - \frac{n_P}{n_S \times n_T}$ is the fraction of time-frequency resources allocated to data.
In conventional systems, increasing pilot density reduces $\rho$, directly lowering goodput even if BLER remains unchanged.
\figref{fig:goodput} shows that, by eliminating pilots without sacrificing reliability, DeepOFDM achieves a substantial goodput improvement over both classical and neural baselines.
Notably, while the standalone NRx with OFDM achieves lower goodput at 100~m/s with one pilot (1P) than two pilots (2P), owing to increased channel aging. DeepOFDM, by contrast, achieves significantly higher goodput by removing pilots altogether, converting the freed time-frequency resources directly into useful data transmission.

\subsection{Comparison to prior pilotless approaches.}
Prior work has explored pilotless or pilot-efficient communication through alternative mechanisms.
Learned geometric shaping for partially coherent reception~\cite{stark2019joint,aoudia2021end} improves robustness to phase uncertainty but retains a one-to-one mapping between symbols and resource elements (GS+NRx).
Another approach is the use of superimposed pilots (SIP) \cite{rezaie2025superimposed}, where pilot energy is embedded directly into data symbols, requiring a neural receiver to jointly disentangle data and reference signals. For each resource element (RE), the transmitted symbol is expressed as \begin{equation} X_{i,k} = \sqrt{1 - A_{i,k}}\,\tilde{X}_{i,k} + \sqrt{A_{i,k}}\,P_{i,k}, \end{equation} where \( \tilde{X}_{i,k} \) denotes the modulated data symbol, \( P_{i,k} \) the pilot signal, and \( A_{i,k} \in [0,1] \) controls the fraction of energy allocated to pilots.

In the fully pilotless setting, DeepOFDM consistently outperforms both GS+NRx and SIP-based pilotless baselines across all evaluated mobility regimes, achieving approximately a $0.5$~dB SNR gain at a target BLER of $10^{-2}$, 100m/s.

Taken together, these results establish pilotless operation as a key capability of DeepOFDM, enabling both reliable communication and improved spectral efficiency in highly dynamic wireless environments.

\section{GENERALIZATION}
\label{sec:generalization}
We next evaluate whether the gains of DeepOFDM persist under changes in OFDM configuration and channel statistics that were not encountered during training.

\begin{figure}[ht]
  \centering
  \includegraphics[width=0.45\textwidth]{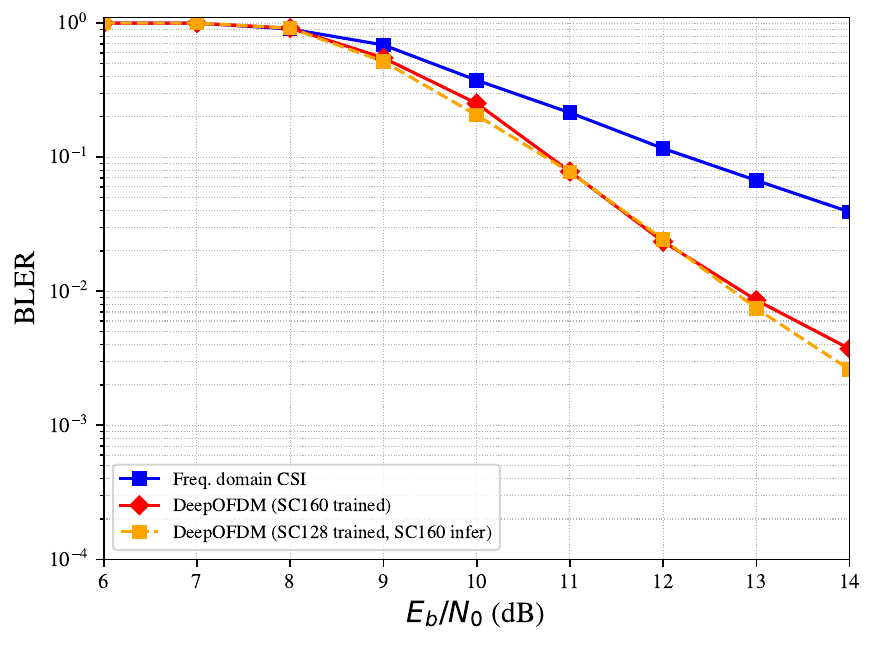}
  \captionsetup{font=small}
  \caption{DeepOFDM trained on a grid with 128 subcarriers transfers to 160 subcarriers without retraining.}
  \label{fig:robustness_sc}
\end{figure}

\subsection{Generalization to unseen channels.}
We next evaluate generalization to channel models that are not encountered during training.
We train DeepOFDM on the TDL-A channel model and evaluate on the CDL-C channel model.
\figref{fig:cdl} reports BLER performance under this train-test mismatch.

Despite the channel mismatch, DeepOFDM maintains its superior performance on CDL-C, approaching the frequency-domain CSI oracle at moderate mobility and outperforming it at high mobility.
Further, DeepOFDM consistently outperforms both QAM+NRx and GS+NRx at moderate and high speeds, mirroring the trends observed on TDL-A.

This indicates that the learned modulation does not overfit to the training channel statistics and transfers to different multipath profiles.
We further evaluate generalization to real-world over-the-air (OTA) channels in \secref{sec:ota}

\subsection{Generalization across OFDM Grid Sizes} We further evaluate DeepOFDM under different OFDM grid configurations by modifying the number of subcarriers and OFDM symbols per slot while keeping the transmitter and receiver architectures fixed.

Specifically, we train the model on an OFDM grid with 128 subcarriers and evaluate it zero-shot on a grid with 160 subcarriers, without any retraining or fine-tuning.

As shown in \figref{fig:robustness_sc}, despite changes in time–frequency resolution, DeepOFDM maintains reliable performance under this mismatch, with only a modest degradation relative to the training configuration.

This behavior can be attributed to the convolutional structure of both the neural modulator and receiver, which operate on local neighborhoods in the time-frequency grid. Since the cyclic prefix structure and subcarrier spacing remain unchanged, the convolutional kernels operate identically on local time-frequency patches regardless of total grid dimension.
As a result, the learned modulation and demodulation rules naturally extend to larger OFDM grids.
In contrast, approaches that rely on explicit grid-specific parameterization such as SIP typically require retraining when the OFDM configuration changes.

\section{OVER-THE-AIR DEMONSTRATION}\label{sec:ota}

While the preceding results demonstrate strong performance in simulation, they rely on statistical channel models. These models capture multipath fading and Doppler dynamics but do not fully represent the impairments present in practical wireless systems. In real deployments, additional effects such as oscillator offset, sampling-rate mismatch, phase noise, and RF hardware nonlinearities can significantly impact receiver performance. To evaluate whether DeepOFDM remains effective under these practical conditions, we deploy the trained models in an over-the-air (OTA) testbed using software-defined radios.

\begin{figure}[ht]
  \centering
  \includegraphics[width=0.45\textwidth]{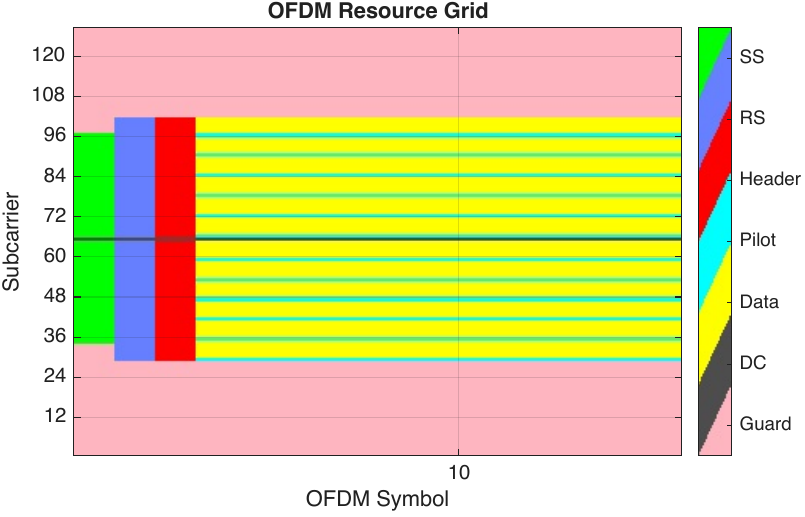}
  \captionsetup{font=small}
  \caption{OFDM frame structure used for OTA transmission, showing synchronization, reference, header, pilot, and data regions used to ensure reliable operation under practical hardware impairments.}
  \label{fig:ofdm_grid}
\end{figure}

\subsection{Experimental Setup}

We implement an over-the-air testbed using two USRP N200 software-defined radios configured in a SISO link. The transmitter and receiver are placed approximately 5 meters apart in an indoor environment, with no mobility. Random data is encoded using a rate $\frac1{2}$ convolutional code and mapped to complex symbols with modulation order $m=6$. The encoded symbols are then processed by the learned neural modulator and passed through the OFDM modulator, which performs a 128-point IFFT and cyclic-prefix insertion to generate the transmit waveform.

\subsection{OFDM Frame}
The transmitted frame structure is illustrated in \figref{fig:ofdm_grid}. Each frame consists of 15 OFDM symbols and includes purpose-built signals to ensure reliable operation: a bandwidth-agnostic \emph{synchronization symbol} (SS) of 31 subcarriers centered at DC for coarse timing/frequency lock; a known \emph{reference symbol} (RS) for initial channel estimation and frequency-offset refinement; a robust \emph{header} conveying bandwidth, subcarrier modulation, and code rate, transmitted immediately after the RS using BPSK at rate~$1/2$ with wide interleaving and a 36-subcarrier DC-centered allocation to maximize detectability under uncertainty; regularly spaced \emph{pilot} tones within the ensuing data region for phase reference and continuous channel tracking (particularly important at higher carrier frequencies); and \emph{DC/guard} nulls to control spectral emissions and suppress LO leakage. The remaining resource elements form the \emph{data region}, where payload symbols occupy the active subcarriers interleaved with pilots, yielding a frame that balances robustness (SS/RS/header/pilots) and spectral efficiency and is used consistently for both the baseline and the proposed DeepOFDM waveform in OTA evaluations.

\subsection{Receiver Processing}

The receiver first compensates for practical channel and hardware impairments such as timing offset, carrier frequency offset (CFO), time-varying fading, and phase jitter. The received signal is first low-pass filtered to suppress out-of-band noise and buffered for synchronization. Frame boundaries are detected by correlating the received samples with the known synchronization symbol, after which automatic frequency control (AFC) estimates and corrects the CFO based on the phase drift observed across the cyclic prefixes of consecutive OFDM symbols. Once synchronization is achieved, the receiver proceeds with OFDM demodulation and reference-symbol-based channel estimation. Two reference symbols from adjacent frames are used to track time-varying fading, with linear interpolation applied to obtain channel estimates for the header and data regions. The header symbol is decoded first to recover system parameters such as FFT size, modulation order, and code rate, enabling proper demodulation of the subsequent data region. Pilot tones embedded within the data symbols are then used to estimate and correct the common phase error (CPE) induced by oscillator drift and phase noise, followed by equalization.

\subsection{Evaluation}
We compare \emph{DeepOFDM+NRx} in pilotless mode against two baselines. First, a neural receiver with learned geometric shaping (GS+NRx) without pilots in the data-region; Second, classical OFDM with two additional pilots for least-squares channel estimation, followed by LMMSE equalization and soft demodulation to LLRs. After equalization, all methods use the same decoding pipeline consisting of deinterleaving, Viterbi decoding, and descrambling prior to CRC verification.

\begin{figure}[ht]
  \centering
  \includegraphics[width=0.45\textwidth]{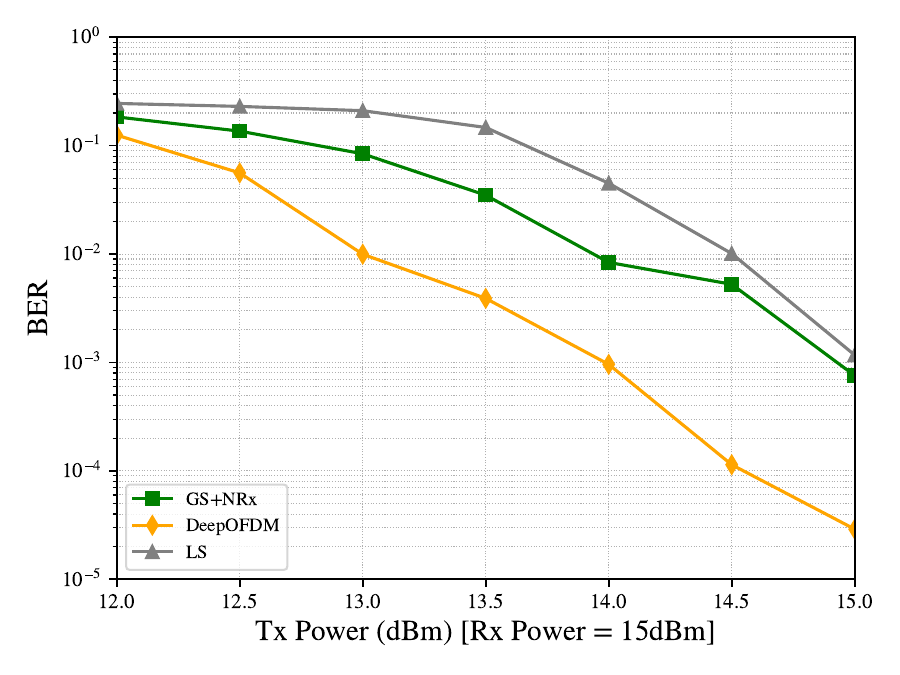}
  \captionsetup{font=small}
  \caption{Over-the-air BER comparison versus transmit power in a static indoor SISO link. Pilotless DeepOFDM achieves target reliability while requiring up to 1 dB lower transmit power than the baselines.}
  \label{fig:ota_ber}
\end{figure}

For a fair comparison, the receiver gain is set to target a received power of \SI{15}{dBm}, while the transmit power is swept from \SI{12}{dBm} to \SI{15}{dBm} in steps of \SI{0.5}{dB}. Since receiver gain is fixed and the noise floor is stable in this controlled environment, transmit power directly corresponds to received SNR, providing a clean single-variable comparison. The resulting BER performance is shown in \figref{fig:ota_ber}. 
Even without pilots, DeepOFDM demonstrates a significant improvement in reliabilit compared to the classical baseline. Compared to GS+NRx, DeepOFDM achieves a given target reliability while requiring up to \SI{1}{dB} lower transmit power, corresponding to approximately $20\%$ power savings in practice.

These results confirm that the gains observed in simulation translate to real wireless environments - DeepOFDM remains robust in the presence of practical hardware impairments such as oscillator offset, sampling-rate mismatch, phase noise, and RF nonlinearities that are not fully captured by idealized channel models.
Finally, the OTA results support the generalization properties observed earlier: the same pretrained DeepOFDM model operates reliably without modification under real channel conditions.

\section{ABLATION STUDIES}\label{sec:ablation}
This section provides detailed ablation studies to understand the key components contributing to DeepOFDM's performance.

\subsection{PAPR Analysis}
Peak-to-Average Power Ratio (PAPR) is a critical metric in OFDM-based systems, as high PAPR can drive power amplifiers into nonlinear regions, leading to spectral regrowth and in-band distortion. 
We evaluate the PAPR characteristics of the proposed DeepOFDM modulation relative to conventional QAM+OFDM.

Empirically, we observe that minimizing the effective rate loss in \eqref{eq:loss} occasionally yields high instantaneous peaks in the time-domain waveform. 
To address this, we introduce a regularization term that penalizes high PAPR during training:
\[
\mathcal{L}_{\text{total}} = \mathcal{L}_{\text{rate}} + \lambda_{\text{PAPR}} \, \mathrm{PAPR}(x),
\]
where $\mathrm{PAPR}(x) = \frac{\max_t |x(t)|^2}{\mathbb{E}[|x(t)|^2]}$ and $\lambda_{\text{PAPR}}$ is gradually annealed toward the later stages of training for stability. 
This annealing schedule prevents premature constraint on the modulator while ensuring that the learned waveform remains power-efficient at convergence.

\begin{figure}[ht]
  \centering
  \includegraphics[width=0.45\textwidth]{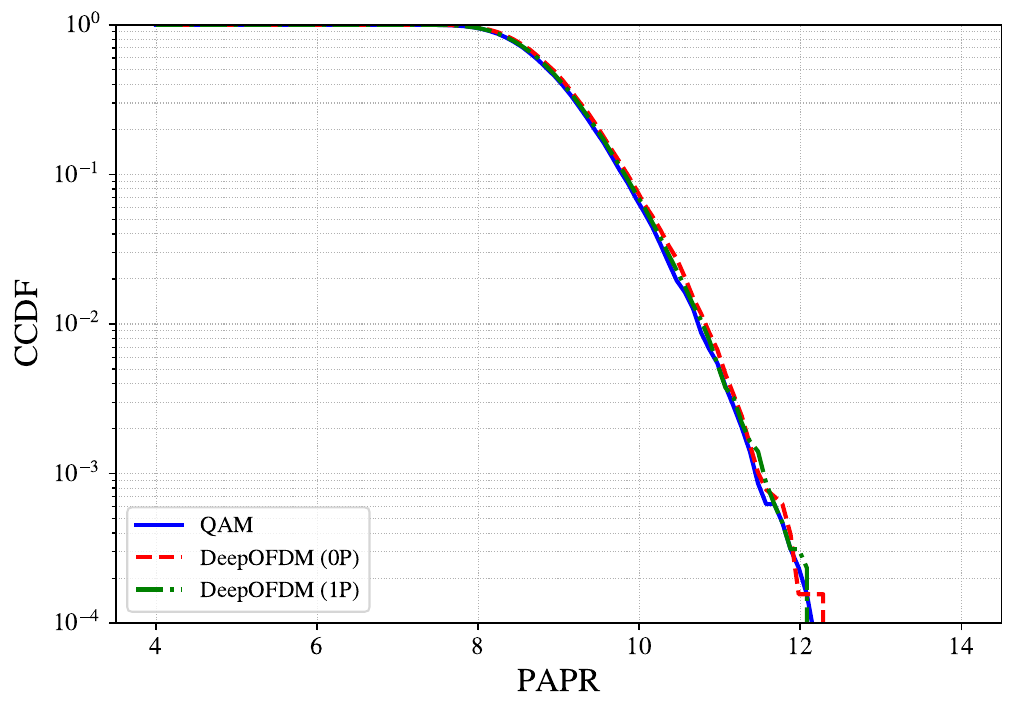}
  \captionsetup{font=small}
  \caption{With the addition of a PAPR regularization term to the loss, DeepOFDM achieves the same PAPR distribution as OFDM}
  \label{fig:papr}
\end{figure}

As shown in \figref{fig:papr}, incorporating this regularization effectively aligns the PAPR of DeepOFDM with that of conventional OFDM using QAM constellations, without degrading BLER performance. 
Thus, DeepOFDM retains the hardware compatibility of standard OFDM transmitters while benefiting from learned modulation robustness.

\subsection{Effect of Nonlinear Time–Frequency Mixing}

We study two additional aspects of DeepOFDM beyond the asymmetric constellation discussed in \secref{sec:interpretation}. In particular, the neural modulator performs time–frequency mixing across the OFDM grid; as a result, the output corresponding to each input symbol does not collapse to a single constellation point, and produces a cluster of outputs depending on the surrounding time–frequency context, as illustrated in \figref{fig:asymeetric}.

\begin{figure}[ht]
  \centering
  \includegraphics[width=0.5\textwidth]{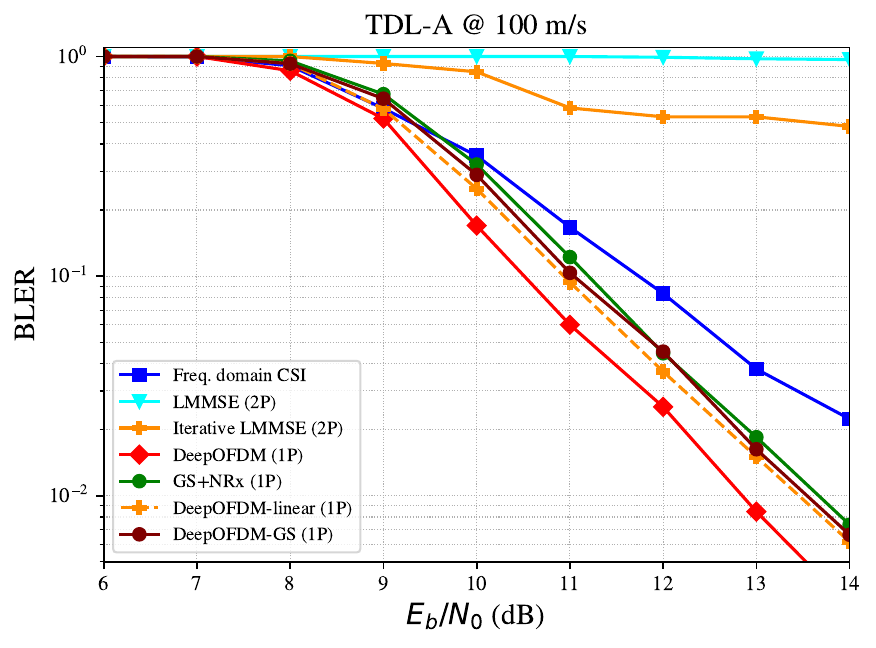}
  \captionsetup{font=small}
  \caption{DeepOFDM-GS collapses the learned outputs into a fixed constellation, while DeepOFDM-linear removes nonlinear activations in the modulator. Both design choices lead to a noticeable degradation, highlighting the utility of non-linear transformations for expressive modulation.}
  \label{fig:linear}
\end{figure}

To isolate the effect of the constellation geometry itself, we construct a baseline denoted \textit{DeepOFDM-GS} by collapsing these outputs into a fixed constellation. Specifically, for each input QAM symbol, we compute the centroid of the corresponding DeepOFDM outputs and treat this centroid as a deterministic constellation point. This produces a geometric shaping scheme derived directly from the trained modulator. As shown in \figref{fig:linear}, DeepOFDM-GS provides only modest gains over GS+NRx, indicating that restricting the learned modulation to a conventional constellation representation removes part of the structure exploited by DeepOFDM.

We next investigate the role of nonlinear transformations in the neural modulator. To this end, we construct a linear variant of DeepOFDM by removing all ReLU activations while retaining the same convolutional layers and batch normalization described in Section~\ref{sec:arch}. This results in a purely linear time–frequency transformation implemented by stacked convolutions. The near-identical performance of DeepOFDM-linear and DeepOFDM-GS suggests that linear time-frequency mixing alone contributes little beyond what fixed constellation shaping already provides; nonlinear activations are what enable the modulator to discover genuinely novel geometric structure, and enables DeepOFDM to achieve substantially lower BLER. 

These results indicate that the gains of DeepOFDM arise from the combination of an asymmetric constellation structure, local time–frequency mixing, and nonlinear transformations within the modulator.

\begin{table*}[ht]
\centering
\small
\setlength{\tabcolsep}{3pt}
\caption{Required SNR (dB) to achieve BLER of $10^{-2}$ across code rates and blocklengths for modulation orders $m=4$ and $m=6$ at 40 m/s.}
\label{tab:deepofdm_results_40ms}
\begin{tabular}{@{}cc|ccc|ccc@{}}
\toprule
\multirow{2}{*}{\textbf{Rate}} & \multirow{2}{*}{\textbf{Blocklength}} & \multicolumn{3}{c|}{\textbf{SNR (dB) @ $m=4$}} & \multicolumn{3}{c}{\textbf{SNR (dB) @ $m=6$}} \\
\cmidrule(lr){3-5} \cmidrule(lr){6-8}
 & & Freq domain CSI & DeepOFDM & GS+NRx & Freq domain CSI & DeepOFDM & GS+NRx \\
\midrule
0.33 & 648 & 5.32 & 6.01 & 6.10 & 7.37 & 7.96 & 9.18 \\
 & 1296 & 4.98 & 5.57 & 6.15 & 7.18 & 7.62 & 8.50 \\
 & 1944 & 4.69 & 5.18 & 6.64 & 6.98 & 7.47 & 8.64 \\
\midrule
0.50 & 648 & 6.59 & 7.28 & 7.37 & 9.42 & 10.01 & 12.60 \\
 & 1296 & 6.35 & 6.98 & 7.32 & 9.29 & 9.77 & 11.13 \\
 & 1944 & 6.25 & 6.69 & 7.71 & 9.13 & 9.57 & 11.72 \\
\midrule
0.67 & 648 & 8.64 & 8.79 & 8.98 & 11.72 & 12.35 & - \\
 & 1296 & 7.91 & 8.69 & 8.79 & 11.43 & 12.21 & - \\
 & 1944 & 7.81 & 8.50 & 9.47 & 11.47 & 12.21 & - \\
\bottomrule
\end{tabular}
\end{table*}

\subsection{Evaluating different code lengths and rates}

The results presented thus far have focused on a single reference configuration: an LDPC code of rate 0.5, with a codelength matched to the maximum number of complex-valued symbols that can fit within one time-frequency resource grid (i.e., a single transport block). This configuration provided a natural baseline for assessing system-level throughput/goodput, as it ensures that the entire OFDM frame is utilized efficiently without the need for code block segmentation or padding.

Now, we assess the generalization and robustness of DeepOFDM across a broader range of coding conditions by evaluating its performance under multiple LDPC code rates, blocklengths, and modulation orders. Specifically, we consider three coding rates:
\begin{itemize}
  \item Low rate: \( R = 1/3 \)
  \item Medium rate: \( R = 1/2 \)
  \item High rate: \( R = 2/3 \)
\end{itemize}

In parallel, we evaluate three blocklength regimes, consistent with those defined in the 3GPP 5G NR LDPC standard:
\begin{itemize}
  \item Short blocklength: \( n = 648 \) bits
  \item Medium blocklength: \( n = 1296 \) bits
  \item Long blocklength: \( n = 1944 \) bits
\end{itemize}

In these settings, multiple independently encoded code blocks are sequentially packed into the same OFDM frame. If the final code block does not completely fill the available time-frequency grid, the remaining resource elements are zero-padded. This structure reflects practical transport block segmentation strategies and enables a consistent comparison of DeepOFDM across different rates and blocklengths, while preserving the same overall frame duration. We also evaluate across modulation orders 4 and 6 - where the input to the neural modulator are complex symbols derived from 16-QAM and 64-QAM respectively.

For each combination of rate and blocklength, we can use the same pretrained DeepOFDM model, since training is code-agnostic, and depends only on the modulation order. In contrast, varying the modulation order (e.g., switching from 16-QAM to 64-QAM) requires retraining a separate model, as the receiver’s learned representation must adapt to the underlying constellation geometry. This code-agnostic behavior is also supported by the OTA experiments in Section~\ref{sec:ota}, where the same DeepOFDM model provides gains when used with a rate-$\frac{1}{2}$ convolutional code instead of LDPC. Performance is evaluated using BLER as a function of SNR, under the same high-Doppler channel conditions described earlier.

As shown in \tabref{tab:deepofdm_results_40ms}, DeepOFDM without pilots consistently outperforms the baseline GS+NRx across all coding rates and blocklengths. The gains are particularly pronounced for higher modulation order $m=6$ - this trend reflects the increasing sensitivity of higher-order constellations to residual channel estimation and phase errors. In contrast, DeepOFDM remains within 
0.4-0.8 dB of the oracle receiver with perfect frequency-domain CSI, indicating that DeepOFDM effectively compensates for a large fraction of the channel estimation error.

Importantly, these gains persist across all evaluated code rates and blocklengths, demonstrating that the benefits of DeepOFDM are largely code-agnostic. Since the neural modulator operates on complex symbols rather than coded bits, the same pretrained model can be used across different LDPC configurations without retraining. This property enables DeepOFDM to integrate seamlessly with standardized coding schemes while maintaining consistent performance improvements across a wide range of operating points.

\subsection{Neural modulation complexity}
\label{sec:complexity}

Our results have demonstrated that redistributing neural network capacity between the transmitter and receiver can significantly increase robustness to Doppler, while keeping the overall model size approximately constant. To better understand this effect, we perform an ablation study that systematically redistributes parameters between the transmitter (Tx) precoder and receiver (Rx) network.

\begin{figure}[ht]
  \centering
  \includegraphics[width=0.5\textwidth]{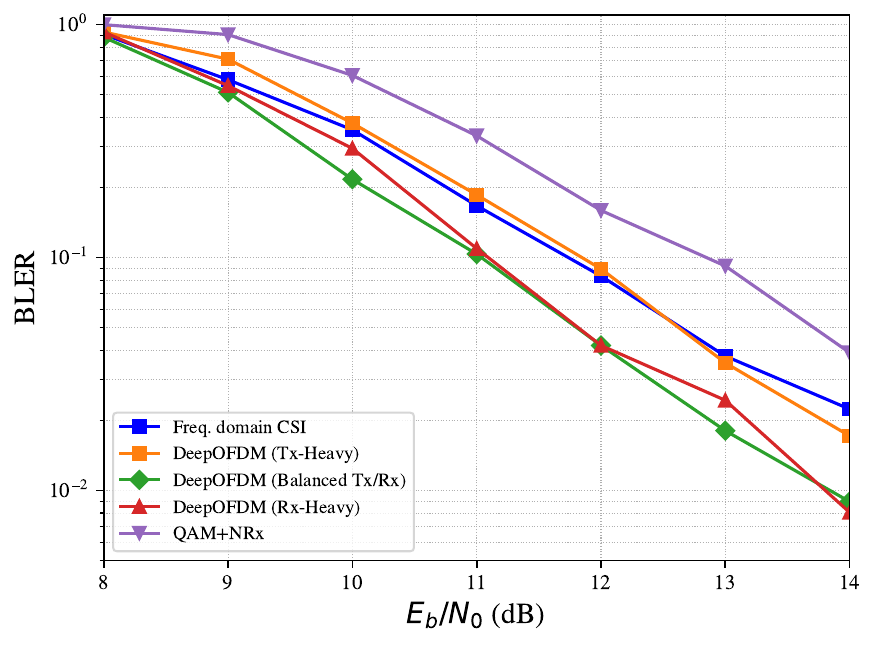}
    \captionsetup{font=small}
  \caption{BLER performance at high mobility ($100$\,m/s) for different Tx/Rx parameter allocations, with similar total parameter budget. Balanced and Rx-heavy configurations achieve strong performance, while Tx-heavy and Rx-only designs degrade due to insufficient receiver capacity for channel estimation and equalization, and lack of sufficient piloting respectively}
  \label{fig:complexity}
\end{figure}

\begin{table}[H]
\centering
\setlength{\tabcolsep}{3.5pt}
\renewcommand{\arraystretch}{1.15}
\small
\captionsetup{font=small}
\caption{FLOPs and parameter counts for Tx/Rx channel-width configurations.
Input: $14$ OFDM symbols $\times 128$ subcarriers; $M{=}6$ bits/symbol}
\label{tab:txrx_complexity}

\begin{adjustbox}{max width=\columnwidth}
\begin{tabular}{l S[table-format=3.2] S[table-format=3.2] S[table-format=3.2]}
\toprule
{Configuration} & {Precoder (MFLOPs)} & {Receiver (MFLOPs)} & {Total (MFLOPs)} \\
\midrule
Rx-Heavy & 65.77  & 446.66 & 512.42 \\
Balanced Tx/Rx & 211.82 & 299.98 & 511.80 \\
Tx-Heavy & 493.05 & 64.94  & 557.99 \\
\bottomrule
\end{tabular}
\end{adjustbox}

\vspace{0.5em}

\begin{adjustbox}{max width=\columnwidth}
\begin{tabular}{l S[table-format=3.2] S[table-format=3.2] S[table-format=3.2]}
\toprule
{Configuration} & {Precoder} & {Receiver} & {Total Params.} \\
\midrule
Rx-Heavy & 18.48  & 124.96 & 143.44 \\
Balanced Tx/Rx & 59.36  & 83.97  & 143.33 \\
Tx-Heavy & 137.99 & 18.23  & 156.22 \\
\bottomrule
\end{tabular}
\end{adjustbox}
\end{table}

We consider three parameter allocation regimes while maintaining a similar total parameter budget. In the \emph{Tx-heavy} configuration, most of the capacity is placed at the transmitter, with channel widths $(48,33)$ for the Tx and Rx networks respectively. The \emph{balanced} configuration distributes parameters evenly between both sides with widths $(40,40)$. Finally, the \emph{Rx-heavy} configuration allocates more capacity to the receiver using widths $(20,50)$.

\tabref{tab:txrx_complexity} summarizes the resulting parameter counts and computational complexity (FLOPs) for these configurations. Although the total parameter budget remains nearly constant, the computational burden shifts significantly between transmitter and receiver depending on the chosen allocation.

\figref{fig:complexity} shows the resulting BLER performance at high mobility ($100$\,m/s), in the single-pilot configuration. Both the balanced and Rx-heavy configurations achieve strong performance, whereas the Tx-heavy and Rx-only designs achieve considerably worse reliability. This asymmetry arises because the receiver must perform channel estimation and equalization regardless of transmitter design. Reducing receiver capacity creates an estimation error floor that transmitter-side learning cannot compensate for.

\section{CONCLUSION AND REMARKS}\label{sec:conclusion}

In this work, we introduced DeepOFDM, a learnable neural modulation scheme that enhances OFDM by replacing fixed constellation mapping with a lightweight CNN-based modulator, jointly optimized with a neural receiver through end-to-end training. 
Our experiments demonstrate that transmitter-receiver co-design provides little gain when the channel is perfectly known, but significantly improves reliability in channel-estimation-limited regimes such as high mobility and sparse pilots. DeepOFDM can also operate without pilots in the data region with minimal performance degradation, improving system goodput. Our findings generalize across unseen channel models and OFDM grid sizes without retraining, and are validated through over-the-air experiments. These results suggest that learned modulation can enhance classical communication systems while remaining compatible with OFDM structure.

Several promising directions remain for future work. First, the architecture naturally extends to MIMO configurations by treating antennas as an additional dimension, and evaluating the interplay between spatial multiplexing and learned constellation asymmetry is a natural next step. Second, while our results establish empirically that breaking rotational symmetry improves phase identifiability under Doppler, a theoretical characterization of how much asymmetry is optimal, and under what channel conditions, remains an open and practically relevant question ~\cite{devroye2022interpreting, zhou2024higher}. Third, further reducing transmitter-side complexity through quantized or pruned modulator designs ~\cite{wei2024advances} would bring DeepOFDM closer to deployment on hardware-constrained platforms.

\vspace{-.8em} 
 \medskip
 \small
 \bibliographystyle{IEEEtran}
 \bibliography{references}

\clearpage
\normalsize

\end{document}